\documentclass[preprint,journal]{vgtc}       % preprint (journal style)

\ifpdf%                                % if we use pdflatex
  \pdfoutput=1\relax                   % create PDFs from pdfLaTeX
  \usepackage{graphicx}                % allow us to embed graphics files
  \DeclareGraphicsExtensions{.pdf,.png,.jpg,.jpeg} % for pdflatex we expect .pdf, .png, or .jpg files
\else%                                 % else we use pure latex
  \usepackage{graphicx}                % allow us to embed graphics files
  \DeclareGraphicsExtensions{.eps}     % for pure latex we expect eps files
\fi%

\graphicspath{{figures/}{pictures/}{images/}{./}} % where to search for the images

\usepackage{microtype}                 % use micro-typography (slightly more compact, better to read)
\PassOptionsToPackage{warn}{textcomp}  % to address font issues with \textrightarrow
\usepackage{textcomp}                  % use better special symbols
\usepackage{mathptmx}                  % use matching math font
\usepackage{times}                     % we use Times as the main font
\usepackage{cite}                      % needed to automatically sort the references
\usepackage{tabu}                      % only used for the table example
\usepackage{booktabs}                  % only used for the table example
\usepackage{subcaption}
\usepackage{float}
\usepackage{url}
\usepackage{diagbox}
\usepackage{amsmath}
\usepackage{multirow}
\usepackage{xcolor}
\usepackage{xspace}
\usepackage{flushend}
\usepackage[normalem]{ulem}
\ieeedoi{10.1109/TVCG.2019.2934557}

\newcommand{\ontoplot}{OntoPlot\xspace}
\newcommand{\protege}{Prot{\'e}g{\'e}\xspace}

\newcommand{\hl}[1]{#1}
\onlineid{1338}

\vgtccategory{Research}
\vgtcpapertype{Application/design study}

\title{\ontoplot: A Novel Visualisation for Non-hierarchical Associations in Large Ontologies}

\author{Ying Yang, Michael Wybrow, Yuan-Fang Li, Tobias Czauderna, Yongqun He}
\authorfooter{
\item
 Ying Yang, Michael Wybrow, Yuan-Fang Li, and Tobias Czauderna are with Monash University, Australia. Email: {ying.yang, michael.wybrow, yuanfang.li, tobias.czauderna}@monash.edu.
\item
 Yongqun He is with University of Michigan Medical School, USA. E-mail: yongqunh@med.umich.edu.
}

\abstract{Ontologies are formal representations of concepts and complex relationships among them. They have been widely used to capture comprehensive domain knowledge in areas such as biology and medicine, where large and complex ontologies can contain hundreds of thousands of concepts. Especially due to the large size of ontologies, visualisation is useful for authoring, exploring and understanding their underlying data. Existing ontology visualisation tools generally focus on the hierarchical structure, giving much less emphasis to non-hierarchical associations. In this paper we present \ontoplot, a novel visualisation specifically designed to facilitate the exploration of all concept associations whilst still showing an ontology's large hierarchical structure. This hybrid visualisation combines icicle plots, visual compression techniques and interactivity, improving space-efficiency and reducing visual structural complexity. 
We conducted a user study with domain experts to evaluate the usability of \ontoplot, comparing it with the de facto ontology editor \protege. The results confirm that \ontoplot attains our design goals for association-related tasks and is strongly favoured by domain experts.}

\CCScatlist{ % not used in journal version
 \CCScat{K.6.1}{Management of Computing and Information Systems}%
{Project and People Management}{Life Cycle};
 \CCScat{K.7.m}{The Computing Profession}{Miscellaneous}{Ethics}
}

\teaser{
  \centering
  \includegraphics[width=\linewidth]{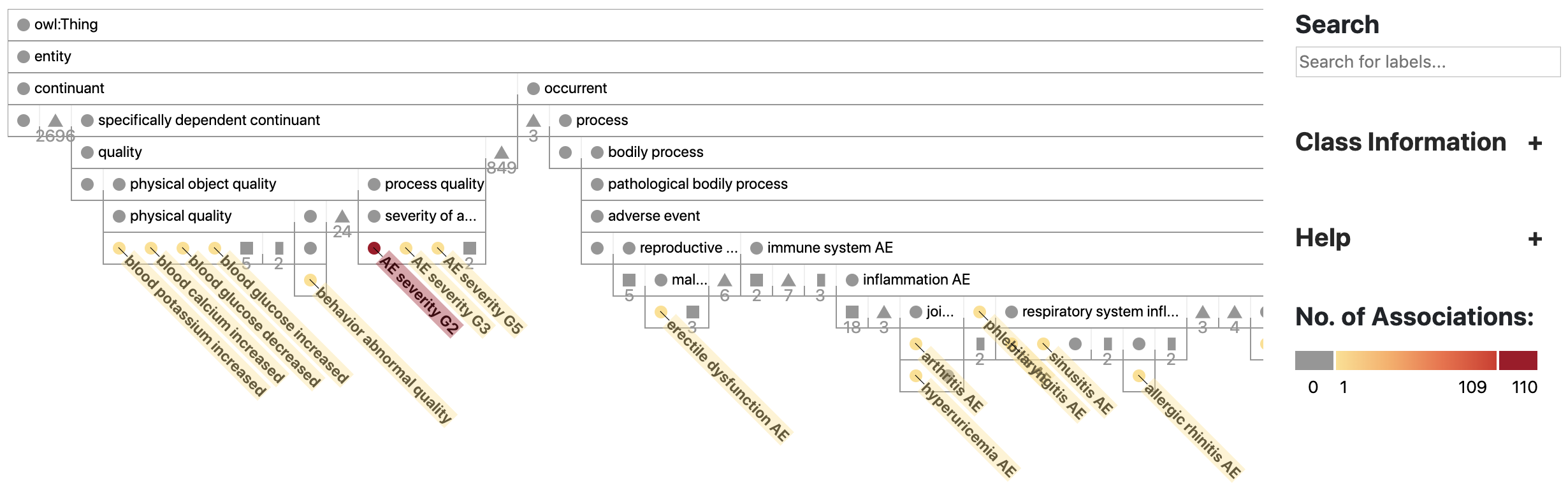}
  \caption{\ontoplot interface. Classes with associations are highlighted in colour based on the key on the right-hand side. A panel on the left (not shown) selects different types of non-hierarchy associations found within the ontology (denoted by the OWL object properties involved). The panel on the right provides search and displays additional information for selected classes (\hl{not shown}).}
  \label{fig:ontoplot-cover}
}

\usepackage{balance}
\begin{document}



\maketitle
%\tcz{Use of subcaption package breaks caption font for figures, tables}

%\tcz{I have set the table to scriptsize as in the template}

%\todo{Consider reviewers' comments from the WebConf submission.}
\section{Introduction}\label{sec:introduction}

%Concept interrelations: direct SubClassOf relationships and other relationships.

%Non-hierarchical associations: relationships contain object property.

%Motivation.

Semantic Web ontologies are widely used in many domains to annotate, retrieve, analyse, and integrate data and knowledge~\cite{citeulike:5008761}. Especially in the biomedical research community, many large and complex biomedical ontologies have been developed to provide a set of formal, controlled and shared 
vocabularies for describing classes and relationships between them~\cite{gene2015gene,shah2009ontologies}. They standardise biomedical concepts and structure their relations with the main aim to support data integration and information exchange~\cite{citeulike:4561539,machado2013semantic}. 

Many large ontologies have been developed over the years. BioPortal~\cite{noy2009bioportal,salvadores2013bioportal}, a comprehensive biomedical ontology repository, currently contains 763 ontologies with a total of almost 10 million classes.\footnote{\url{https://bioportal.bioontology.org/}} These include the influential Gene Ontology~\cite{gene2015gene}, which has close to 50,000 classes, the SNOMED CT ontology~\cite{stearns2001snomed}, which currently has more than 340,000 classes, among others. Visual tool support for the effective interrogation of these ontologies is essential to both ontology users and ontology developers. As a response to this important problem, many visualisation systems have been developed in recent decades~\cite{katifori2007ontology,saghafi2016visualizing,dudas2018}. 

Visualisation maps information to a graphical representation to convey knowledge. Effective visualisation makes it possible to obtain insights into data, support user tasks, and perform exploration on large and complex data structures. It typically reduces cognitive effort by limiting the amount of information presented to users.

Ontologies, especially those expressed in the OWL~\cite{hoph03a} and OWL 2 languages~\cite{Grau:2008:ONS:1464505.1464604}, usually have a hierarchical, tree-like structure defined by subsumption relationships between concepts (or classes\footnote{In the rest of this paper we use the terms concept and class interchangeably.}). Given the central importance of subsumption relationships in defining an ontology, many of existing visualisation systems rightly treat hierarchies as first-class citizens. For example, Figure~\ref{fig:protege-configuration} shows the interface of \protege~\cite{noy2000knowledge}, the de facto ontology editor. The left-hand side of the interface shows the subsumption hierarchy of the ontology in an indented tree layout. 

Besides the subsumption relationship between \emph{named} classes, important \emph{associations} between classes are also expressed through subsumption and the use of properties and anonymous class expressions (cf Axiom~\ref{axm:assc}). These associations between concepts in biomedical ontologies define the relational expressions between the concepts in the biomedical domain~\cite{smith2005relations} and they capture a variety of rich information in addition to the subsumption hierarchy. 

% \yy{The associations in biomedical ontologies can be seen as solution providers for experimental analysis. They define the relational expressions between the concepts in the biomedical domain~\cite{smith2005relations}.} \yf{remove this para.}

\hl{For example, anatomy ontologies~\cite{sprague2006zebrafish,mungall2012uberon,Hayamizu2015MouseAO} model the parts of organisms and the structural and developmental relationships between these parts. The rich set of spatial associations in anatomy ontologies can be used to define overlapping, continuous, or adjacent regions. Thus, it is important to see which parts of an organism are spatially associated with another part of that organism. Also, the regulatory associations in the Gene Ontology indicate where one process or function affects the manifestation of another process, function, or quality~\cite{gene2009gene}. Multiple molecular functions regulate one target, reflecting cooperative translational control, while one molecular function may have multiple targets, indicating target multiplicity~\cite{enright2003microrna,yoon2005prediction,nam2007mirgator,liu2012identifying}.} %RNA regulation papers

\hl{Adverse drug events (ADEs) are undesired medical consequences of drug interventions~\cite{he2014oae,zaman2017use}. According to~\cite{winnenburg2015exploring,zaman2017use}, ADEs result in more than 770,000 injuries and deaths each year and cost up to \$5.6 million per hospital, and can lead to withdrawal of marketed drugs or failure of drug development. Therefore, identifying and predicting ADEs are major focuses in pharmacovigilance. Many ontologies have been developed to capture and analyse ADEs~\cite{kuhn2010side,lin2012ontology,iyer2013mining,he2014oae,sarntivijai2016linking,guo2016ontology,wang2017ontology,hur2018ontology}. In the ontological context, ADEs are often investigated on the class level, where a given ADE may be common to all drugs in the corresponding class, or conversely, an ADE may be associated with some class members but not with all of them. The discovery of such associations facilitates the learning and prediction of ADEs. It is also important to see which drug associates the most adverse events and vice versa.}

%\tcz{I'm not sure about the next two sentences. Combine into one sentence? Reverse order? It's quite a jump from an explanation what associations are to an example of negative side effects of a drug.} \yf{Revised the example.} 

% \yy{%For example, the ODNAE ontology~\cite{guo2016ontology} models drugs' \emph{adverse neuropathy events}, where a drug may contain chemical compounds that can have negative side effects on the nervous system. %For example, the drug Bupropion Oral Tablet contains a certain chemical compound (bupropion) and (indirectly) negatively regulates neurotransmitter uptake. In the ontology, they are represented as a number of subsumption \emph{axioms} and displayed in separate views in \protege, as can be seen in Figure~\ref{fig:protege-default}.
% More than 20 distinct neuropathy adverse events have been defined in ODNAE, and a number of drugs have been found to be associated each of these adverse events~\cite{guo2016ontology}. 
%ODNAE and similar ontologies provide important domain knowledge for identifying \emph{drug class effects}, which are identical or similar effects of a group of drugs, an important task in pharmacovigilance.
% } 

\hl{Despite the importance of associations, most of the existing ontology visualisation tools focus on class hierarchy~\cite{katifori2007ontology,saghafi2016visualizing,dudas2018}. Although some tools, such as~\cite{storey2001jambalaya,jurcik2012knoocks,da2012integrated,kuhar2012ontology,lohmann2014webvowl}, do visualise ontology associations, they do not support the complex association tasks mentioned above, and users sometimes need to count the number of associations between multiple classes to perform such tasks. Additionally, existing ontology visualisation tools are often vague about the use cases and tasks they support~\cite{carpendale2014ontologies,dudas2018}.} 

% \yy{%However, existing hierarchy-centric tools such as \protege do not facilitate the discovery of complex association tasks mentioned above, and a user needs to count the number of associations between multiple   combinations of drugs and adverse effects to determine this~\cite{guo2016ontology,wang2017ontology}. % are not easy to visualise, which presents challenges to the understanding of ontologies.
% }

In this paper, we present \ontoplot (available at \url{https://ialab.it.monash.edu/ontoplot/}), a novel ontology visualisation system specifically designed to support association-related user tasks. It includes a number of novel features:
\begin{itemize}

\item It uses a hybrid visualisation combining icicle plots~\cite{kruskal1983} with visual compression techniques to show an ontology's inheritance backbone.

\item \hl{It automatically compresses irrelevant subtrees to effectively emphasise non-hierarchy associations} and uses \hl{distinct} glyphs to help distinguish different \hl{structures} of compressed subtrees.

% \item It uses a visualisation similar to an icicle plot~\cite{kruskal1983} with compression to show an ontology's subsumption backbone.

\item \hl{It allows users to interactively expand and collapse subtrees of the hierarchy, with additional functions like filtering by a particular property or class, search, and highlighting by colours and labels.}

% \yy{%\item It supports intuitive search, filtering by properties, highlighting non-hierarchy associations by colours and labels, and automatic collapsing of irrelevant subtrees to effectively emphasise associations.
% }

% \item It supports intuitive search, highlighting, filtering by properties, and automatic collapsing of irrelevant subtrees to effectively show non-hierarchy associations. 
\end{itemize}

To evaluate \ontoplot's effectiveness in supporting user tasks, we conducted a prototype evaluation and then a subsequent user study with domain experts. 
% \hl{using a large ontology of adverse events of cardiovascular drugs}. 
Results show that for association-oriented tasks, \ontoplot outperforms \protege in terms of accuracy and task completion time.

\hl{%\sout{Biomedicine domain dominates the coverage of ontology use, since bioinformatics has developed and used more ontologies than other fields~\cite{d2012publish}.} 
While ontologies are most widely used in the biomedical domain~\cite{d2012publish}, similar tasks are performed with ontologies in a number of application areas. For example, in the agronomy domain, many ontologies have been produced to represent and analyse agronomic data~\cite{jonquet2018agroportal,drury2019survey}. They are used to answer questions like ``what are the appropriate rice varieties for a given soil or region?'' and ``how many rice varieties are bred from a particular breeding station?''. In the e-government domain, they are used to analyse the information about citizens, authorities, or investment~\cite{fraser2003knowledge,wagner2006building}. For bibliometrics, they are used to investigate the research areas, scientific collaborations, publication impact factor, and granted funding of researchers, their affiliations and regions, leading to potential opportunities~\cite{adam2002citation,moed2006citation,peroni2018spar}. \ontoplot is domain agnostic and can facilitate equivalent tasks for any domain.}
% We performed a user study to evaluate \ontoplot's effectiveness in supporting user tasks. Results show that for complex association-oriented tasks, \ontoplot outperforms \protege in terms of accuracy, task completion time, learning effort and user preference.

\section{Related Work}\label{sec:related}

Here we give an overview of biomedical ontologies and discuss biomedical ontology associations \hl{as examples of use cases to which \ontoplot can be applied}, describe visualisation methods for ontologies, and then explore approaches for visual compression.

\subsection{Ontology}\label{sec:ont}

The term \textit{ontology} comes from philosophy. It refers to the study of things that exist in nature and how to describe and group them. It has been adopted by computer science to represent a formal specification of conceptualisation~\cite{Antoniou:2012:SWP:2381011}. 

An ontology describes knowledge in a domain. It consists of concepts and relationships between these concepts. Typically, a concept is a set of objects, and their relationships are defined as binary relations. The most prevalent relationship is that of \emph{inheritance}, where one class $C$ is stated as a \emph{subclass} of another class $C'$, if every object in $C$ is included in $C'$. The inheritance relationships in an ontology typically form a tree-like hierarchy. %For example, in a university domain, staff, academic staff, and administration staff are some concepts. The hierarchical relationships between these concepts are: academic staff is a subclass of staff; and administration staff is a subclass of staff.

For example, the Cardiovascular Disease Ontology (CVDO)~\cite{barton2014cardiovascular} is a medium-sized ontology with approx.\ 500 concepts. The concept `familial atrial fibrillation' (ID \verb+DOID_0050650+) is a subclass of concept `atrial fibrillation (disease)' (ID \verb+CVDO_0000092+). 

Besides the inheritance hierarchy, ontologies also define other types of relationships between classes. These relationships capture richer associations between classes within the same ontology or even across different ontologies. These associations are typically defined over \emph{predicates} (binary relations) and other concepts. 

In CVDO, the concept `familial atrial fibrillation' is further constrained such that it must be mapped through the predicate `has material basis at all times' (ID \verb+BFO_0000113+) to another concept, `genetic disorder' (ID \verb+OGMS_0000047+).

Expressed in the OWL DL syntax~\cite{hoph03a}, the above two definitions can be formally expressed as follows. Axiom~\ref{axm:subc} states the subclass relationship (denoted `$\sqsubseteq$') between `familial atrial fibrillation' and `atrial fibrillation (disease)'. Axiom~\ref{axm:assc} states the association on `familial atrial fibrillation', asserting it as a subclass of a \emph{someValuesFrom} value restriction. 

\begin{align}
\texttt{DOID\_0050650} &\sqsubseteq \texttt{CVDO\_0000092}\label{axm:subc}\\
\texttt{DOID\_0050650} &\sqsubseteq \exists\ \texttt{BFO\_0000113}.\texttt{OGMS\_0000047}\label{axm:assc}
\end{align}

% The class hierarchical relationships in an ontology are not required to form a strict hierarchy, meaning that a class may have multiple superclasses. More specifically speaking, if a class $A$ is a subclass of both $B_1$ and $B_2$, then every object of $A$ is an object of $B_1$ and $B_2$. Ontologies are structured in such a way that there could be multiple paths to a term in the hierarchy. 

Ontologies have been widely adopted for the purpose of knowledge representation in a number of areas, especially in biological and medical research. A key motivation of their adoption is that an ontology can provide a basis for integrating and understanding knowledge from multiple sources. In this research, we will only discuss biomedical ontologies and their use. However, we note that \ontoplot is not restricted to only this domain, \hl{as described in Section \ref{sec:introduction}.} 

% \subsubsection{Biomedical Ontologies}\label{sub:biomedical_ontology_associations}

% \tcz{I'm not sure if we need a subsubsection here. There is not a second one.}

Biomedical research is one of the popular applications of ontologies~\cite{stevens2009application,shah2009ontologies}, where they drive the computational use of biological data. In biomedical research, there is an abundance of heterogeneous data, including genes, proteins, clinical observations, and laboratory data, that need to be integrated to facilitate the formulation, evaluation, and refinement of hypotheses. Biomedical ontologies achieve this by organising and classifying knowledge in a formalised and structured manner, providing unambiguous and shareable descriptions. 

The shared understanding of collected data is essential for biologists to describe the same entities in the same way. One typical example is the widely-used Gene Ontology~\cite{ashburner2000gene}, which defines a large number of concepts (or classes or terms) to annotate biological entities (i.\,e., genes and gene products) that result from high-throughput experiments.%To tackle this challenge, controlled vocabulary is delivered as the primary and the most common used artefact of ontology in bioinformatics. A controlled vocabulary is a list of specifically predefined terms.  

%Typically, biomedical ontologies offer the measure of semantic similarity between two biological entities. For example, the Gene Ontology structure itself is widely used for genes semantic similarity analysis. 
%
% \subsubsection{Biomedical Ontology Associations}
%
% Biomedical ontologies can also be designed to serve as a reference source to support analysis purposes. The created ontology is used to integrate the terms from different ontologies, and provide a link to analyse the relationships between these ontologies. In this case, the relationships are mainly many-to-many. 
%

% \yy{TODO: explain class effect here.}

One example ontology with rich non-hierarchical associations is the Ontology of Drug Neuropathy Adverse Events (ODNAE)~\cite{guo2016ontology}. ODNAE is developed to support the study of drug-associated neuropathy adverse events. It extracts classes from different ontologies and integrates them to generate an ontology-based semantic framework that brings all related knowledge together in a logical and structured format for interdisciplinary representation and analysis. Extending the Ontology of Adverse Events (OAE)~\cite{he2014oae}, ODNAE imports related drugs from the Drug Ontology (DrON)~\cite{hanna2013building} with their chemical components defined in the Chemical Entities of Biological Interest (ChEBI) ontology~\cite{hastings2012chebi}, drug mechanisms of action from NDF-RT \cite{brown2004va}, and biological process in the Gene Ontology (GO)~\cite{ashburner2000gene}, and links these classes with semantic relations. Totally, ODNAE contains 1,579 classes. 

While performing analysis of drug-associated neuropathy adverse events, queries can be used on ODNAE to answer specific questions, such as: ``how many associations between drugs and their corresponding neuropathy adverse events are at different levels in the hierarchy?'', ``how many neuropathy-inducing drug chemicals are classified at different levels of ChEBI?'', and ``how many adverse events are related to different groups of drug molecular entities?''. 
\hl{One significant question is about the \emph{class effect}. Given an adverse event and a drug class, a class effect exists for the drug class when all its subclass drugs (drug chemical ingredients or drug products) are associated with the adverse event.
In other words, if there is a class effect~\cite{winnenburg2015exploring}, it means that the effect is exhibited by every subclass of the class.}
Non-hierarchical associations are essential in answering these queries, and we believe an intuitive, task-supportive visualisation can assist people to better understand such complex ontologies.

\subsection{Ontology Visualisation}\label{sec:ont_vis}
With the increased adoption of ontologies in diverse fields, there is a growing need for effective ontology visualisations to support development, management, and utilisation of ontologies.

Compared to visualising strict hierarchies, ontologies are more challenging. Firstly, ontologies often contain multiple inheritance. %that rules out a whole class of visualisation techniques. 
This is typically solved by duplicating a concept under each of its parents or by using multiple edges to link a concept to all of its parents, either of which has its own drawbacks. With duplication there is the problem of redundancy, whereas with multiple edges there is the problem of visual occlusion. 

Secondly, an ontology can contain a rich set of non-hierarchical relationships, which are typically defined using object properties, datatype properties, or annotation properties. Hierarchies are also used to represent other types of information, including concept equivalence and disjointness~\cite{DBLP:conf/dlog/BaaderN03}. %Concepts and relations in an ontology can also have attributes, such as value restriction, or disjointness restriction. 
However, most ontology visualisations still target the hierarchical structure of ontologies. Some of them visualise all the relations, while some visualise exclusively the hierarchy. Moreover, each concept may have instances, ranging from one or two to thousands~\cite{katifori2007ontology}. Depending on the task, sometimes instances are required to be visualised.

%\tcz{Any references for the first part of this paragraph?} \yf{Added one.}

% A few visualisations have been developed to focus on modelling all the relations in ontologies via visual notations. Figure \ref{fig:visual_notations} shows two famous examples. As this research addresses the hierarchical structure of ontologies, the visualisation of visual notations will not be discussed in detail. 

The most widely used ontology visualisation method is that of indented trees, which is employed by the de facto ontology editor \protege~\cite{noy2000knowledge}. It primarily visualises the inheritance hierarchy of an ontology and duplicates concepts for multiple inheritance. Non-hierarchical associations are listed textually in a separate pane. 
%The default interface of \protege can be seen in Figure~\ref{fig:protege-default}. 

\hl{Network diagrams are another popular method to visualise ontologies. OWLViz~\cite{horridge2005owlviz}, a plugin for \protege, uses a layered node-link (network) diagram to visualise the inheritance relationships, providing an alternative view for the hierarchy, but does not display any non-hierarchical associations. WebVOWL~\cite{lohmann2014webvowl} was developed as a visual notation for OWL. It models concept interrelations in ontologies but does not visually differentiate hierarchical relationships and non-hierarchical associations.} 

%A plugin OWLViz~\cite{horridge2005owlviz} serves as an alternative basic visualisation for \protege, providing a layered node-link diagram for the ontology inheritance hierarchy, but this suffers from scalability issues when the ontology is big.

\hl{Several well-known tools such as Jambalaya~\cite{storey2001jambalaya}, Knoocks~\cite{jurcik2012knoocks}, OntoViewer~\cite{da2012integrated}, and a multiple view visualisation tool developed by Kuhar and Podgorelec~\cite{kuhar2012ontology} use node-link or space-filling strategies to represent the ontology inheritance hierarchy structure, and visualise the non-hierarchical associations as links between the classes in the hierarchy. None of these tools appear to be actively maintained.}

%Several other well-known plugins for \protege such as Jambalaya~\cite{storey2001jambalaya}, and Knoocks~\cite{jurcik2012knoocks} were developed to support the display of the ontology inheritance hierarchy in a hierarchical structure, and both use a separate graph to visualise non-hierarchical associations. However, these plugins are no longer maintained. 
% to the best of our knowledge none of these tools are available any more and are no longer maintained.

%More recently, VOWL 2~\cite{lohmann2014vowl} was developed as a visual notation for OWL. It uses graph-based visualisation to model concept interrelations in ontologies, but does not visually differentiate hierarchical relationships and non-hierarchical associations. WebVOWL~\cite{lohmann2014webvowl} is an online visualisation system that implements the visual notation defined in VOWL 2. In our preliminary study we encountered scalability issues with WebVOWL when visualising medium to large ontologies (hundreds or thousands of classes and their associations). 

A comprehensive survey on ontology visualisation~\cite{katifori2007ontology} categorises systems for visualising ontologies based on their visualisation types: indented list, node-link and tree, zoomable, space-filling, focus + context or distortion, and 3D information landscapes. A recent survey~\cite{saghafi2016visualizing} proposes two categories: graph-based methods and multi-method visualisation techniques. The latest \hl{survey}~\cite{dudas2018} \hl{provides a useful classification and comprehensive evaluation of available ontology visualisation tools. The results} show that most visualisation systems focus on class hierarchies, and that \hl{their} maturity, usability, \hl{and scalability} are still limited.
%This research groups ontology visualisation tools based on different visual channels mentioned above: connection, containment, adjacency, and multiple views, which employ different channels in separate views. 
Interested readers are referred to these surveys for a comprehensive overview of ontology visualisation methods.

\subsection{Visual Compression of Large Ontologies}

The visualisation of large ontologies, or large hierarchies in general, is challenging.  The difficulty for ontology developers during the creation of an ontology is how to use available screen space for presentation most effectively. For ontology users the challenge is how to explore and interact with a large ontology most efficiently.

The visualisation of large ontologies, or large hierarchies, can be done using explicit methods (explicit representation of parent-child relations, e.g., by edges) or implicit methods (implicit representation of parent-child relations, e.g., by positional encoding). The later of the two approaches allows for four axes in the design space: 1) dimensionality (2D or 3D), 2) node representation (graphics, primitives, glyphs), 3) edge representation (inclusion, overlap, adjacency), and 4) layout (subdivision, packing)~\cite{schulz2011design}.

In particular, the utilisation of glyphs for the representation of groups of nodes for the visualisation of graphs and hierarchies has drawn some attention in recent years since it allows for a more compact representation by compressing or simplifying the topology. 
\hl{For graphs,} Dunne and Shneiderman introduced motif simplification for node-link diagrams which replaces common patterns of nodes and links with compact and meaningful glyphs~\cite{dunne2013motif}. 
%However, of their motifs---fans of nodes with a single neighbour, connectors that link anchor nodes, and cliques of completely connected nodes---only the fan motif seems to be relevant for ontology visualisation. 
A similar approach to motif simplification has been proposed by Shi et al.~\cite{Shi2013}. Their  structural equivalence grouping considers nodes with similar connectivity behaviour and patterns (but not necessarily close proximity) as a group. 
However, this approach seems to be limited to a particular topology. Yoghourdjian et al. proposed graph thumbnails for identification and comparison of large graphs~\cite{Yoghourdjian2018}, but these visual summaries \hl{hide a lot of connectivity information} necessary to understand associations.
More comprehensive discussions of glyph-based visualisation strategies, guidelines and techniques in general can be found in~\cite{Ward2002} and~\cite{Borgo2013}.

For hierarchies, the Cheops method uses triangles to visually compress hierarchical datasets based on context and user interaction but is limited to horizontal compression~\cite{Beaudoin1996}. \hl{Jiao et al.~\cite{jiao2013visualization} apply the compression technique to leaf nodes if the number of leaves for a parent node is above a certain threshold, and use a single large node to represent those leaves to save space.} Heer and Card~\cite{heer-avi-2004} have explored Degree-of-Interest (DOI) trees where some uninteresting branches are collapsed so the tree can be arranged within a constrained area. They allow users to interactively explore by collapsing and expanding, and they progressively recompute DOI values.  More recently,  Nobre et al.~\cite{Nobre2018} used similar hierarchical DOI compression for family trees. Their visualisation displays a hierarchy arranged horizontally with a separate row dedicated to each person of interest, each displaying \hl{multiple} attributes that can be easily compared between \hl{nodes}. They summarise uninteresting subtrees and siblings by showing them as small icons on the rows of interest. 
%While each of these approaches are effective for their individual aims, both of them come at the cost of making the hierarchy more difficult to understand.

Approaches for the visual compression of large hierarchies have, to the best of our knowledge, not been applied to the visualisation of large ontologies and their non-hierarchical associations. However, it seems that these approaches are very suitable to create more compact visualisations and to make it easier to interactively explore such visualisations. %The aforementioned indented tree visualisation of ontologies provides some compression but with a focus on hierarchy, without consideration of non-hierarchical associations.
\hl{While glyph encoding techniques can convey abstract structural information and hide complexity~\cite{munzner2003treejuxtaposer}, they could be improved to give a visual summary of hidden structure.}

\section{\ontoplot Design}\label{sec:design}

In this section we list the use cases for biomedical ontologies that were our original motivation, identify design requirements arising from these, and then describe the \ontoplot visualisation in detail.

\subsection{Motivation} % (fold)
\label{sub:design_motivation}

% \begin{table*}[!htb]
% \caption{Common use cases when working with association data in biomedical ontologies.}
% \label{table:use_cases}
% \scriptsize%
% \centering%
% \begin{tabu}{lll}
% \toprule
%  Label & Description & Need\\
% \midrule
% U1 & Discover new knowledge & Access the entire ontology and its contained information.\\
% U2 & Generalise concepts & See the path from a class to the root.\\
% U3 & Discover common knowledge & Find the lowest level of common ancestors for associations.\\
% U4 & \hl{Explore a class' associations} & See the distribution as well as details of associations.\\
% U5 & Detect significant associations & Compare relative association strength of classes.\\
% U6 & Identify class effect & See when associations apply to a number of child classes.\\
% U7 & Predict possible associations & Show the sibling of classes with associations.\\
% \bottomrule
% \end{tabu}
% \end{table*}

\begin{table}[!htb]
\caption{Common use cases when working with association data in biomedical ontologies.}
\label{table:use_cases}
\scriptsize%
\centering%
\begin{tabu}{lll}
\toprule
 Label & Description & Need\\
\midrule
U1 & Discover new knowledge & Access the entire ontology and\\
& & its contained information.\\
U2 & Generalise concepts & See the path from a class to the\\
& & root.\\
U3 & Discover common knowledge & Find the lowest level of com-\\
& & mon ancestors for associations.\\
U4 & \hl{Explore a class' associations} & See the distribution as well as\\
& & details of associations.\\
U5 & Detect significant associations & Compare relative association\\
& & strength of classes.\\
U6 & Identify class effect & See when associations apply to\\
& & a number of child classes.\\
U7 & Predict possible associations & Show the sibling of classes\\
& & with associations.\\
\bottomrule
\end{tabu}
\end{table}

In Section~\ref{sec:ont} we outlined the process of using biomedical ontologies for exploring and cataloguing the adverse effects from drug use. 
%One of our co-authors has significant expertise using biomedical ontologies for exploring and cataloguing the adverse effects from drug use, a process outlined in Section~\ref{sec:ont}. 
This type of work involves understanding not only the underlying hierarchy, but also types and strengths of associations between classes in different parts of the ontology. \hl{During early  interviews with our co-author, Yongqun He, an expert in bioinformatics,} we identified a number of common use cases that are important for such work (see also Table~\ref{table:use_cases}):

% \begin{table}[!htb]
% \centering
% \caption{Common use cases when working with association data in biomedical ontologies.}
% \label{table:use_cases}
% \begin{tabular}{ l | >{\arraybackslash}p{1in} | >{\arraybackslash}p{1.5in}}
%  \textbf{Label} & \textbf{Description} & \textbf{Need}\\
% \hline
% U1 & Generalise concepts & See the path from a class to the root\\
% U2 & Discover common knowledge & Find the lowest level of common ancestors for associations\\
% U3 & Detect significant associations & Compare relative association strength of classes\\
% U4 & Identify class effect & See when associations apply to a number of child classes\\
% U5 & Predict possible associations & Show the siblings of associations\\
% \end{tabular}
% \end{table}

% \begin{figure*}[!htb]
%   \centering
%   \includegraphics[width=0.98\textwidth]{ontoplot-cover}

%   \caption{\ontoplot interface. Classes with associations are highlighted in colour based on the key on the right-hand side. The panel on the left selects different types of non-hierarchy associations found in the ontology (denoted by the OWL object properties involved) . The panel on the right provides search and displays additional information for selected classes.}
%   \label{fig:ontoplot-basic}
% \end{figure*}

\begin{itemize}
\item U1: In order to discover new knowledge, users need to be able to access the entire ontology and its contained information.
\item U2: In order to generalise concepts, users must be able to clearly trace the path from a given concept to the root concept of the hierarchy.
\item U3: In order to discover common knowledge, given two classes of interest the user must be easily able to trace their paths towards the root and determine the concept that is the lowest common ancestor.
\item U4: \hl{For a class of interest,} users must be able to see the distribution of associations \hl{across} the ontology hierarchy. They must also be able to see the \hl{number of associations and association details}.
\item U5: In order to detect significant associations, users must be able to easily identify the associations with the greatest strength in the ontology, and the classes to which they apply.
\item U6: In order to identify class effects \hl{(as described in Section~\ref{sec:ont})}, users need to be able to clearly identify when particular types of associations apply to most or all child classes of a given class, i.e., the association effects on a class of things.
\item U7: In order to predict possible associations, users need to be able to explore the siblings of classes with a given association.
%\yf{Is it children or siblings as in Table~\ref{table:use_cases}?} 
%\mjw{Ying had ``siblings of associations'' originally.  I assume this is because siblings could be expected to have same associations as each other.}
\end{itemize}

\subsection{Design Requirements} % (fold)
\label{sub:design_requirements}

From the use cases described above, along with the typical nature of ontological data, we can identify a number of design requirements for an interactive system to explore associations.

\hl{As described in U1, the entire ontology should be embodied in the visualisation.} Since ontologies can be very large, \hl{the visualisation is required to maximise the use of available space (R1).}

\hl{Also, ontologies are} generally broad (i.e., much wider than they are deep), %we require the visualisation make use of available screen space as much as possible.  As many computer monitors and laptop screens have a wide-screen aspect ratio, this means there is a need to maximise the use of horizontal space.
with traditional tree visualisations, the branches with large numbers of leaf nodes will take up significant amounts of horizontal space. There is a need to give less prominence to branches with large numbers of leaf nodes. \hl{Moreover, to support the access of information contained in large ontologies, users should be able to explore the ontology and easily find desired information (R2). }

Ontologies encode a clear hierarchical structure through sub-class-of relationships.  Even though we are primarily interested in non-hierarchical associations, the hierarchy is still the most useful way to arrange ontologies, so this must be prominently represented \hl{(R3). In addition, the hierarchical structure is essential for use cases U2 and U3.}

% yy: copied to previous paragraph
%An issue with traditional tree visualisations is that branches with large numbers of leaf nodes take up significant amounts of horizontal space.  Ontologies often have very large numbers of leaf nodes, making their visualisations very wide.  There is a need to give less prominence to branches with large numbers of leaf nodes.

When considering associations in large ontologies, those associations might only apply to a small subset of the ontology.  An effective visualisation needs to clearly highlight the parts of the ontology with relevant associations \hl{(R4) and emphasise those with the greatest strength (U5).}
% When associations relate to all children of a particular class, then class effect needs to be clear. %\mjw{I realise the \ontoplot isn't really supporting class effect well.}

Furthermore, where there are large parts of the ontology without relevant associations, the visualisation should be able to hide or show these \hl{(R5)} so that the user can consider just the relevant parts of the ontology \hl{(U4, U6, U7)}, or optionally view additional information \hl{(R6)} as desired \hl{(U4)}.

% subsection design_requirements (end)

\subsection{Visual Design} % (fold)
\label{sub:visual_representation_design}

% \begin{figure}[!htb]
%   \centering
%   \subfloat[Leaf nodes, expanded.\label{fig:compression-leaves-expanded}]{\includegraphics[width=0.4\textwidth]{compression-leaves-expanded}}
%   \hfill
%   \subfloat[Leaf nodes, collapsed.\label{fig:compression-leaves-collapsed}]{\includegraphics[width=0.4\textwidth]{compression-leaves-collapsed}}

%   \subfloat[Chain, expanded.\label{fig:compression-chain-expanded}]{\includegraphics[width=0.4\textwidth]{compression-chain-expanded}}
%   \hfill
%   \subfloat[Chain, collapsed.\label{fig:compression-chain-collapsed}]{\includegraphics[width=0.4\textwidth]{compression-chain-collapsed}}
    
%   \subfloat[Subtree, expanded.\label{fig:compression-subtree-expanded}]{\includegraphics[width=0.4\textwidth]{compression-subtree-expanded}}
%   \hfill
%   \subfloat[Subtree, collapsed.\label{fig:compression-subtree-collapsed}]{\includegraphics[width=0.4\textwidth]{compression-subtree-collapsed}}

%   \caption{Examples of \ontoplot visual compression.}
%   \label{fig:compression}
%   \todo{Recreate the figures with new OntoPlot}
% \end{figure}

%\begin{figure}[!htb]
%  \centering
%  \includegraphics[width=0.35\columnwidth]{ontoplot-key}
%
%  \caption{\ontoplot colours for associations. Unique colours show the minimum and maximum number of associations, and remaining colours show the range between these.}
%  \label{fig:ontoplot-key}
%\end{figure}

% Here we describe the design of the \ontoplot system, and discuss how it satisfies the described use cases and design requirements.

\ontoplot is similar in style to an icicle plot~\cite{kruskal1983}.  The basic visual style of \ontoplot can be seen in Figure~\ref{fig:ontoplot-cover}. It primarily emphasises the tree structure of the hierarchy using boxes \hl{(R3)}, where the children of a given item are displayed directly below the parent item, and the parent box's width is the total width of all of its children.  However, the use of a standard icicle plot to visualise an ontology hierarchy with many leaf nodes would not be ideal since the overall width of the visualisation would be proportional to the number of leaf nodes. For this reason, we take the basic layout of an icicle plot but represent nodes in the hierarchy as circle glyphs within the boxes traditionally used in icicle plots. Where a number of a given node's children are leaf nodes, we consolidate \hl{(wrap)} these together in a single box that is taller and wider in order to accommodate multiple circle glyphs (see Figure~\ref{fig:compression-leaves-expanded} and~\ref{fig:compression-subtree-expanded}).  Thus, we reduce the overall width of the visualisation at the cost of a moderate increase in height \hl{(R1)}. \hl{A similar approach of wrapping leaves was identified in~\cite{graham2007visual} which arranges leaf nodes in grids under the enclosure of their parent node.}
Even with this \hl{wrapping} of leaf nodes, a large ontology may still be very wide.  For this reason, \ontoplot lets the user easily scroll the visualisation horizontally.

\begin{figure}[!htb]
  \centering
  \includegraphics[width=0.50\columnwidth]{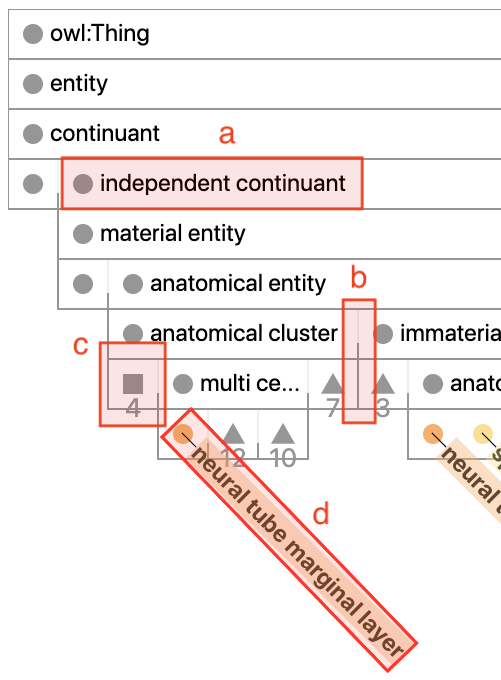}

  \caption{\ontoplot visual design. (a) Parent classes are labelled where possible. (b) Solid lines are shown between different subtree boxes (bottom) and a partial, faint line is shown between sibling boxes (top). (c) Uninteresting subtrees are compressed.  (d) Classes with associations are labelled and coloured based on the number of associations.}
  \label{fig:ontoplot-features}
\end{figure}

\ontoplot does not display labels for all classes by default since these take up a large amount of space and can cause problems with occlusion when densely packed. Boxes for displayed subtrees are often quite wide so labels for these nodes are shown greedily where space exists (see Figure~\ref{fig:ontoplot-features}a).  To differentiate between neighbouring boxes that contain siblings of the same parent class versus neighbouring boxes from different subtrees, \ontoplot uses either a partial and faint line in the first case and a solid line in the second case (see Figure~\ref{fig:ontoplot-features}b).

When \ontoplot is loaded, the user is presented a list of non-hierarchy association types (predicates as described in Section~\ref{sec:ont}) found within the ontology.  They select the association type they are interested in (with the option to change this at any time). \ontoplot visualises these associations by labelling and colouring the circle glyphs of classes to which that type of associations apply.  These classes are frequently, but not always, leaf nodes in the hierarchy. A range of colours are used, where intensity is used to indicate the number of associations applying to that class.  The colour key is dynamic depending on the maximum number of those associations applying to any one class.  Nodes with the minimum and maximum number of associations are clearly coloured \hl{(R4)} and further colours are used to categorise values interpolated between these (e.g., see bottom-right in Figure~\ref{fig:ontoplot-cover}).  This colouring is also applied to the labels for these classes to make the associations stand out. Labels for classes with associations are positioned diagonally to allow labelling of neighbouring classes without occlusion. 

When the user selects an association type, much of the hierarchy will be \textit{uninteresting} in the sense it doesn't contain any classes with these associations.  \ontoplot detects these and uses a form of visual compression to collapse uninteresting subtrees within the ontology (see Figure~\ref{fig:ontoplot-features}c).  This is described in detail in the following section.

% subsection visual_representation_design (end)

\subsection{Visual Compression} % (fold)
\label{sub:visual_compression_design}

While considering a particular type of associations and the classes to which they apply, there will often be a large subset of the ontology which is \emph{uninteresting} in terms of those associations. For this reason, we designed a form of visual compression that allows us to compress the uninteresting subtrees in order to give more prominence to the \emph{interesting} parts of the hierarchy \hl{(R5)}.
 
\begin{figure}[!htb]
  \centering
  \begin{subfigure}[b]{0.45\columnwidth}
    \centering
    \includegraphics[width=0.7\columnwidth]{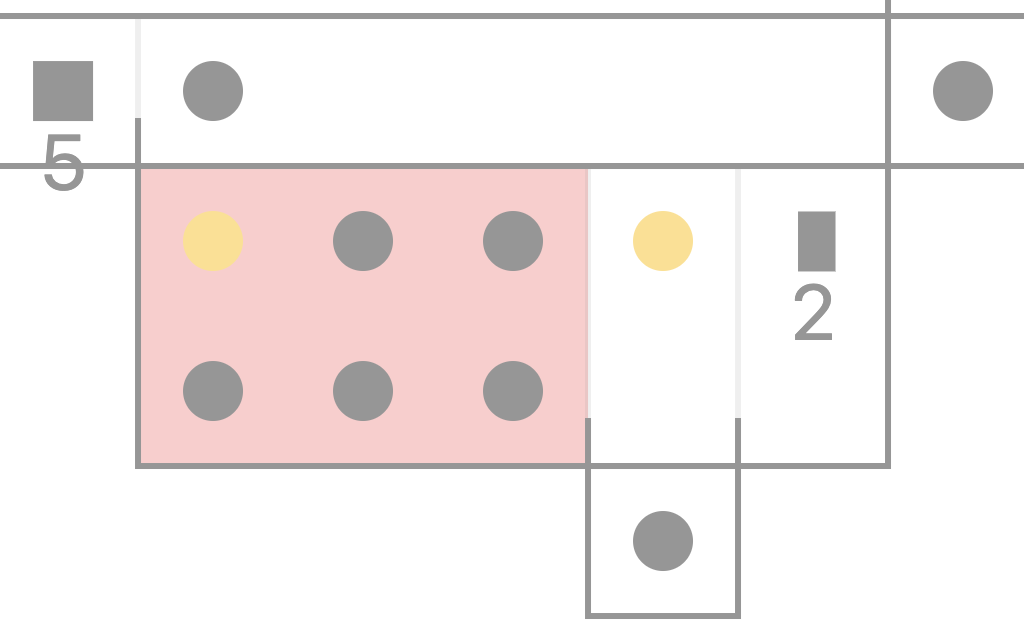}
    \caption{Leaf nodes, expanded.}
    \label{fig:compression-leaves-expanded}
  \end{subfigure}
  \hfill
  \begin{subfigure}[b]{0.45\columnwidth}
    \centering
    \includegraphics[width=0.7\columnwidth]{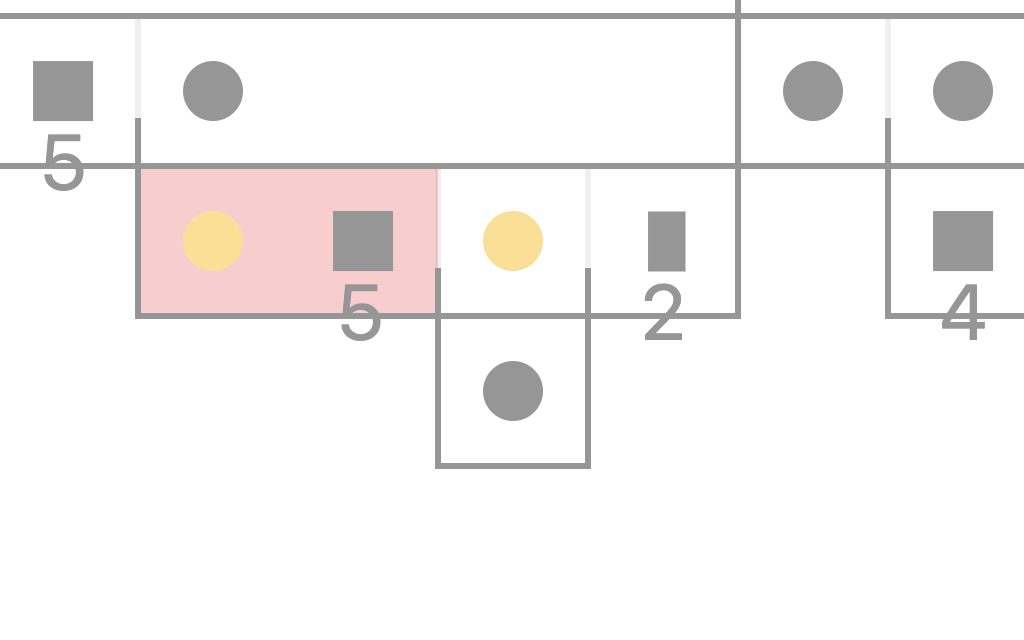}
    \caption{Leaf nodes, collapsed.}
    \label{fig:compression-leaves-collapsed}
  \end{subfigure}

  \begin{subfigure}[b]{0.45\columnwidth}
    \centering
    \includegraphics[width=0.82\textwidth]{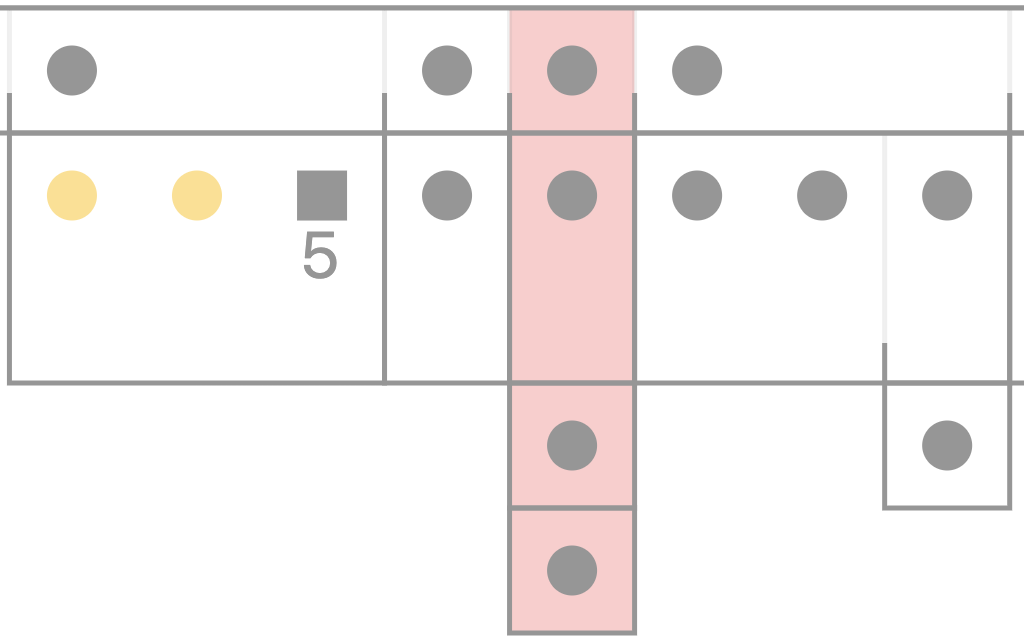}
    \caption{Chain, expanded.}
    \label{fig:compression-chain-expanded}
  \end{subfigure}
  \hfill
  \begin{subfigure}[b]{0.45\columnwidth}
    \centering
    \includegraphics[width=0.82\textwidth]{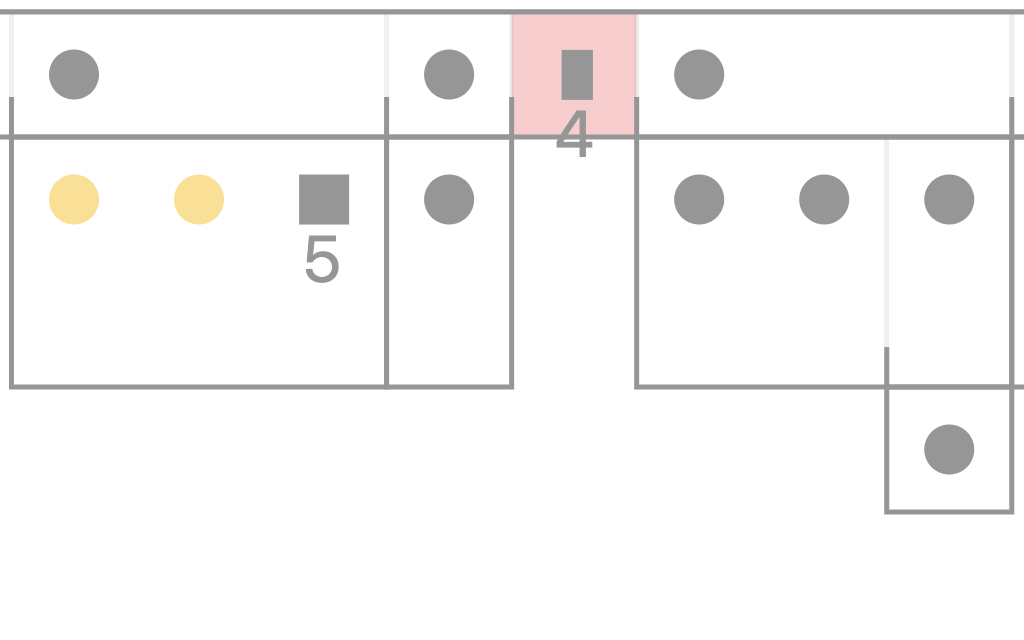}
    \caption{Chain, collapsed.}
    \label{fig:compression-chain-collapsed}
  \end{subfigure}

  \begin{subfigure}[b]{0.45\columnwidth}
    \includegraphics[width=\textwidth]{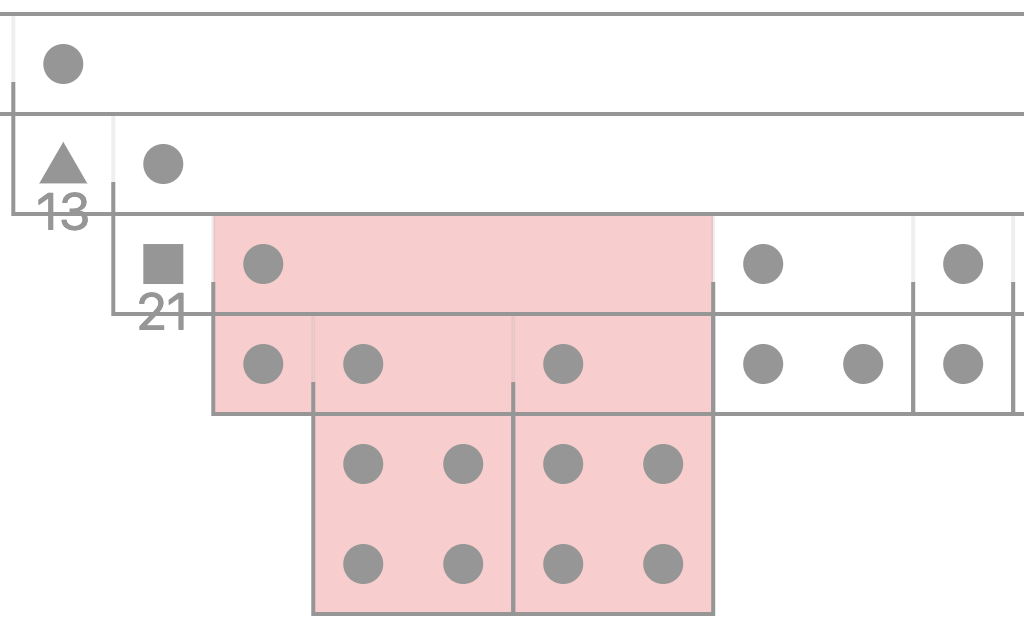}
    \caption{Subtree, expanded.}
    \label{fig:compression-subtree-expanded}
  \end{subfigure}
  \hfill
  \begin{subfigure}[b]{0.45\columnwidth}
    \includegraphics[width=\textwidth]{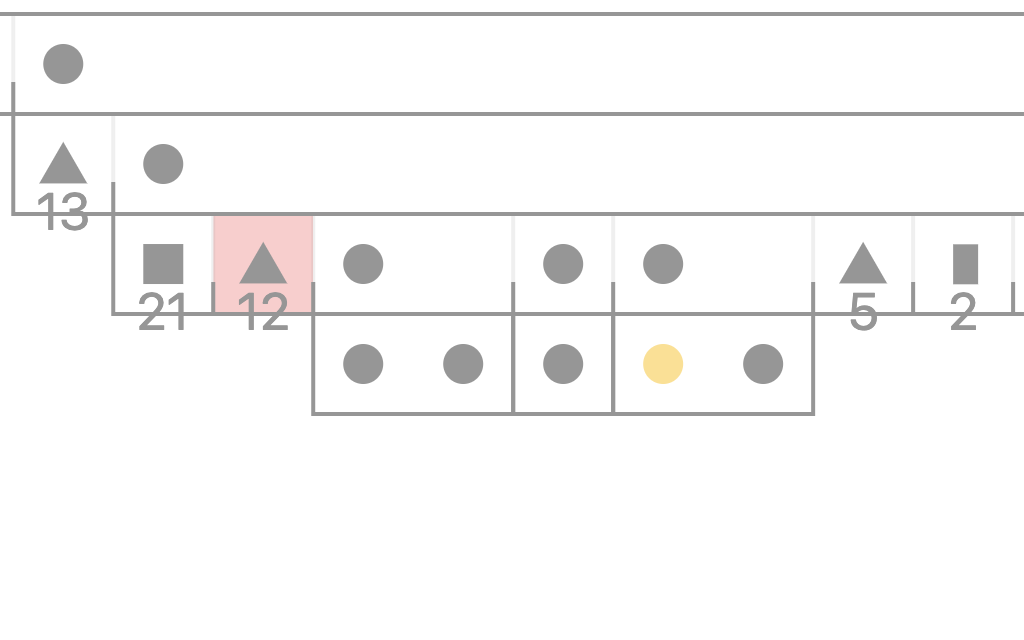}
    \caption{Subtree, collapsed.}
    \label{fig:compression-subtree-collapsed}
  \end{subfigure}

  \caption{Examples of \ontoplot visual compression.}
  \label{fig:compression}
\end{figure}

We identified three cases worthy of compression \hl{and use distinct glyphs to represent their different structure} (see Figure~\ref{fig:compression}):
\begin{itemize}
\item \textbf{Leaf nodes}: Where an interesting node has multiple uninteresting leaf nodes as children, these nodes will already be shown as a number of circle glyphs in a single box.  We replace these multiple dots with a single square glyph, labelled with the number of hidden nodes.  
\item \textbf{Chain}: Where an interesting node has a descendent subtree that is a chain of only uninteresting nodes, we replace this chain with a single box containing a thin block glyph, labelled with the number of nodes in the chain.
\item \textbf{Subtree}: Where an interesting node has a descendent subtree that contains only uninteresting nodes and doesn't fall into the previous two categories, we replace the subtree with a single box containing a triangle glyph, labelled with the number of nodes in the subtree.
\end{itemize}

Whenever we show a particular set of associations, each class in the hierarchy can be considered \emph{interesting} or \emph{uninteresting} depending if it has any associations.  Using this, we can walk the hierarchy to get the set of boxes that can be compressed as described above.  We do this by performing a recursive depth-first traversal of the tree from the root box, where for each box the recursive call returns the number of interesting nodes in the subtree and an array of any collapsible boxes.  If the active box is \emph{interesting}, it adds to the collapsible box array any of the children that contain no interesting nodes.  Since we can discover these collapsible nodes in a single depth-first traversal, this process is linear in the number of classes. 

% subsection visual_compression_design (end)

\subsection{Interaction} % (fold)
\label{sub:interaction}

\begin{figure}[!htb]
  \centering
  \begin{subfigure}[b]{0.48\columnwidth}
    \centering
    \includegraphics[width=0.7\columnwidth]{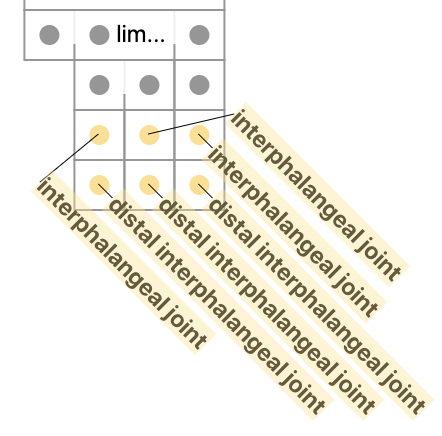}
    \caption{Association labels for classes are initially positioned diagonally below the class they label to minimise overlaps, but can be manually dragged \hl{and arranged} by the user\hl{, as shown}.}
    \label{fig:ontoplot-drag-label}
  \end{subfigure}
  \hfill
  \begin{subfigure}[b]{0.5\columnwidth}
    \centering
    \includegraphics[width=0.7\columnwidth]{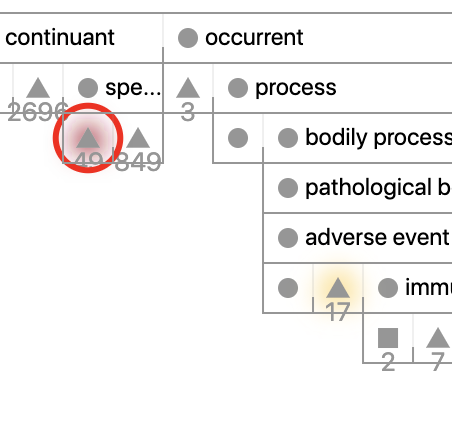}
    \caption{The collapsed subtree containing association classes and the selected class will be highlighted with a coloured shadow (for the maximum number of associations) and a pulsing red circle.}
    \label{fig:ontoplot-collapse}
  \end{subfigure}
  
  \caption{Interactive features of \ontoplot.}
\end{figure}

\begin{figure}[!htb]
  \centering
  \includegraphics[width=0.70\columnwidth]{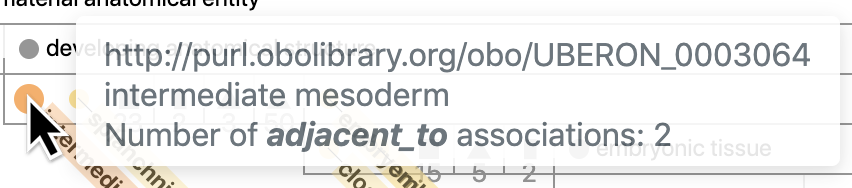}

  \caption{\ontoplot shows labels while hovering over classes.}.
  \label{fig:ontoplot-label}
\end{figure}

\ontoplot displays labels for parent nodes greedily where possible, and labels for classes with associations as described earlier. The user can click and drag these associated class labels if they happen to be obscured (shown in Figure~\ref{fig:ontoplot-drag-label}). For all other classes, \ontoplot displays class labels and other information in a pop-up window while the user hovers over the glyph corresponding to the class \hl{(R2)}, as shown in Figure~\ref{fig:ontoplot-label}.

% \begin{figure}[!htb]
%   \centering
%   \includegraphics[width=0.30\columnwidth]{ontoplot-drag-label}

%   \caption{\yy{\ontoplot shows association labels that have been dragged to remove the overlapping.}}.
%   \label{fig:ontoplot-drag-label}
% \end{figure}

While \ontoplot performs automatic visual compression to hide the uninteresting parts of the ontology hierarchy, the user can always interactively expand and collapse subtrees in order to show or hide sections of the ontology. To expand a subtree, the user can double-click on the glyph of a compressed section of the tree (square, thin block or triangle).  The user can also compress a particular subtree by double-clicking on the glyph corresponding to the root of that subtree \hl{(R2)}. Subtrees can be collapsed regardless of whether the classes they contain are interesting or uninteresting. If a subtree contains interesting classes, the glyph is displayed with a coloured shadow to which indicates the maximum number of \hl{associations} in the collapsed subtree (shown in Figure~\ref{fig:ontoplot-collapse}).

% \begin{figure}[!htb]
%   \centering
%   \includegraphics[width=0.90\columnwidth]{ontoplot-collapse}
%   \caption{\yy{When collapsing a subtree containing association classes and the selected class, the glyph of this subtree will be highlighted with a coloured shadow (for the association classes) and a pulse red circle (for the selected class) in \ontoplot.}}. 
%   \mjw{Ying, can you crop this one, and put Figure 4 and this one in the same figure as subfigures, to save space.}
%   \label{fig:ontoplot-collapse}
% \end{figure}

Interactive collapsing or expanding operations are performed efficiently without recomputing and redrawing the entire visualisation. Instead we compute just the changes (box size and box position translations) that need to be performed to sections of the visualisation as the result of any interaction.  To help preserve the user's mental map, \ontoplot highlights the portion of the tree being collapsed or expanded prior to the operation and then highlights the same \hl{portion} for a moment after the operation has completed.  This highlighting is shown in Figure~\ref{fig:compression}.

\begin{figure}[!htb]
  \centering
  \includegraphics[width=0.70\columnwidth]{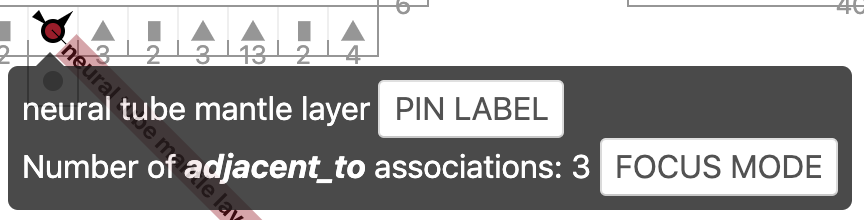}

  \caption{When a class is selected in \ontoplot highlights it and shows additional information and controls.}.
  \label{fig:ontoplot-selection}
\end{figure}

The user can select or deselect a class by clicking on it.  When they do so the visualisation updates to show label and highlight with colour only the classes that the selected class has associations with.  The selected class is given a black outline and subtle arrows denoting the direction of the associations (one pointing in at the top-left if it is the target of the selected association type and another pointing out at the top-right if it is the source of the selected association type).  While a class is selected, the right-hand panel of the interface displays additional information on that class, including a textual list of all its associations \hl{(R6)}. A popover window is also displayed below the selected node (see Figure~\ref{fig:ontoplot-selection}). This indicates the number of associations and gives the user a ``Pin Label'' button to mark the class with its label and a ``Focus Mode'' button \hl{(R5)} to compress the hierarchy to show only associations the selected class is involved in (described below).
If the user collapses a subtree containing \hl{a} selected class, a pulsing red circle will be shown around the glyph for the collapsed subtree (shown in Figure~\ref{fig:ontoplot-collapse}).

The right-hand panel also offers a search field.  When the user enters a search term, that panel displays a scrollable list of matching classes in the ontology.  Selecting a class from this list selects the class in the visualisation and scrolls the main \ontoplot view to make that class visible \hl{(R2)}.  If the user ever scrolls the visualisation away from the selected node, \ontoplot shows a pulsing arrow at the edge of the view that points to the glyph for the selected class.

\subsection{Focus Mode}
\label{sub:focusmode}

\begin{figure}[!htb]
  \centering
  \includegraphics[width=0.9\columnwidth]{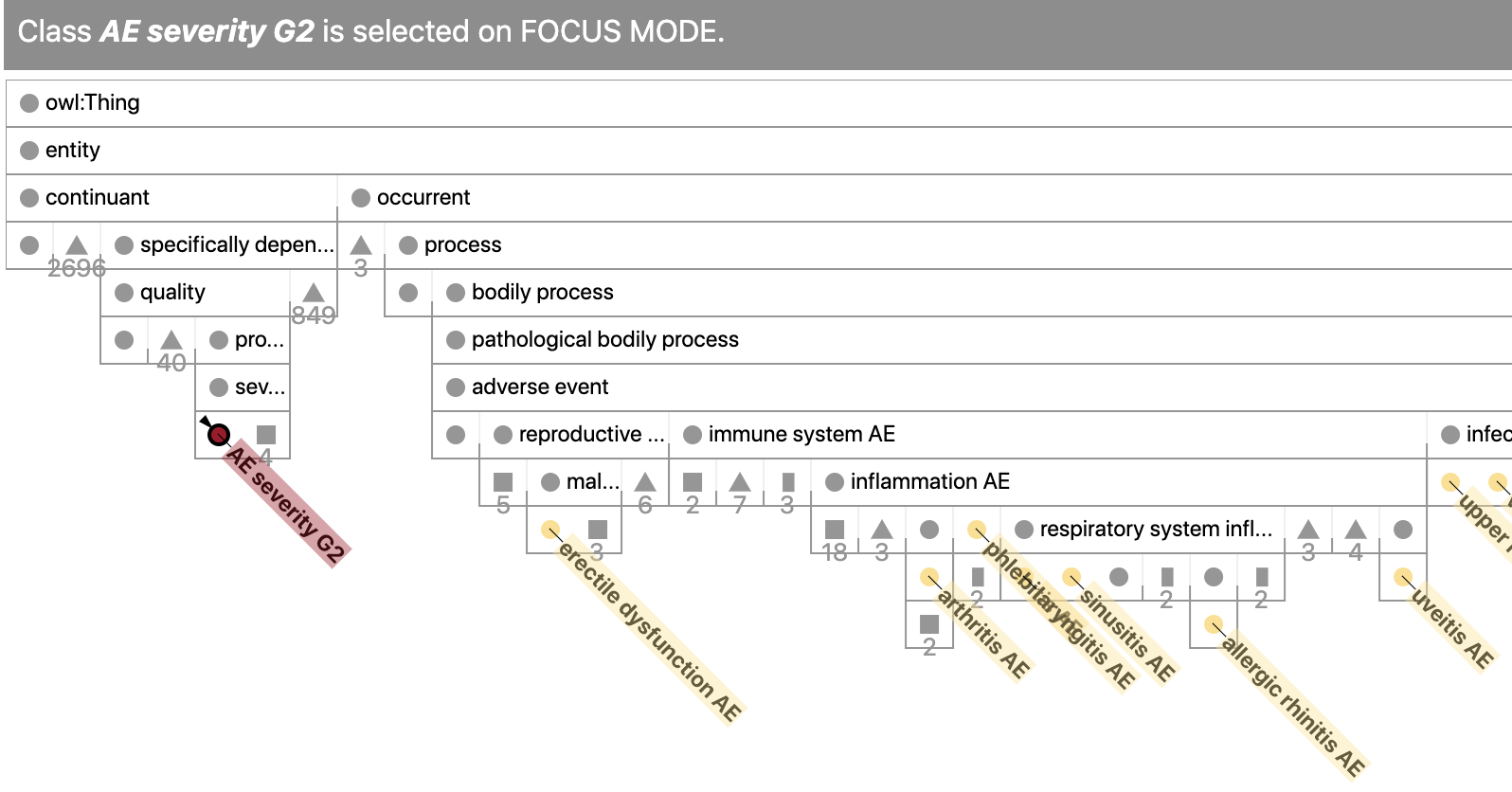}

  \caption{\ontoplot showing the same ontology as in Figure~\ref{fig:ontoplot-cover} but after the user has selected \textit{focus mode} for class `AE severity G2' to concentrate on the associations only for this class.}.
  \label{fig:ontoplot-focus}
\end{figure}

When the user wants to focus on the associations for a particular class.  They can click the button shown in the popover below the selected node.  This causes \ontoplot to recompute the interesting and uninteresting parts of the ontology and visually compress the uninteresting subtrees to show a view that emphasises just the selected class and other classes directly associated with that class (as shown in Figure~\ref{fig:ontoplot-focus}).  The user can interactively explore this view including expanding and collapsing subtrees.  While in this mode, a dark bar is shown at the top of the interface as a reminder.  A ``Reset View'' button is available that returns to a view of the hierarchy which shows all classes with the given association type. 

% \begin{figure}[!htb]
%  \centering
%  \includegraphics[width=0.98\columnwidth]{ontoplot-focus}
%
%  \caption{\ontoplot showing the same ontology as in Fig.~\ref{fig:ontoplot-cover} but after the user has selected \textit{focus mode} to concentrate on the associations selecting a single class. \yy{This image is with a different property in Fig.~\ref{fig:ontoplot-cover}. I did a screen shot of the focus mode with the same property as shown below~\ref{fig:ontoplot-focus_v2}. Please feel free to choose either one.}}.
%  \label{fig:ontoplot-focus}
%\end{figure}

% subsection interaction (end)

%\input{user_study}

%\input{improvement.tex}

\section{Prototype Evaluation}\label{sec:user_study_1}

We conducted a user-based evaluation of an earlier prototype version of \ontoplot. We recruited 20 participants, including 2 domain experts and 18 general users for this controlled experiment. The study design and procedure were similar to the expert user study described below.  
Participants performed tasks with two different sized ontologies to gauge performance at two distinct levels of difficulty; CVDO~\cite{barton2014cardiovascular} (536 classes) and OCVDAE~\cite{wang2017ontology} (4,589 classes).
Overall, the results showed that \protege~\cite{noy2000knowledge} slightly outperformed \ontoplot for most of the Hierarchy-related tasks on both accuracy and completion time. For Association-related tasks, \ontoplot significantly outperformed \protege on accuracy, but the completion times using both tools were similar. The detailed results for accuracy, completion time and subjective rating of this study can be found in the supplementary materials.
% \footnote{Available at \url{http://ialab.it.monash.edu/ontoplot/study/}.}  
We don't present details of the prototype study in this paper for several reasons; firstly, the study identified a number of issues with the \ontoplot interface which we have since addressed (again, see details in the supplementary materials), secondly, the study design had some issues (the training was inadequate and not every participant completed each task for all ontologies), and thirdly, most participants were non-expert users (which makes the results less appropriate for evaluating our original aims). 

\section{Expert User Evaluation}\label{sec:user_study_2}

In order to determine if the design of the \ontoplot system meets our original design requirements, we conducted an expert user study with 12 \hl{new} participants\hl{, all} domain experts or experienced ontology users.

\subsection{Study Design}
\label{sub:study_design}

In the user study, we compare \ontoplot with \protege~\cite{noy2000knowledge}. 

\hl{As mentioned in Section~\ref{sec:ont_vis}, several tools  support the display of non-hierarchical associations alongside an ontology's inheritance hierarchy. When considering these for our preliminary study, we encountered scalability issues with WebVOWL~\cite{lohmann2014webvowl} when visualising medium to large ontologies (hundreds or thousands of classes and their associations). Jambalaya~\cite{storey2001jambalaya} and Knoocks~\cite{jurcik2012knoocks} are no longer maintained and do not run. Neither OntoViewer~\cite{da2012integrated} nor the multiple view tool described in~\cite{kuhar2012ontology} are publicly available.}

\begin{figure}[!htb]
  \centering
  \includegraphics[width=\columnwidth]{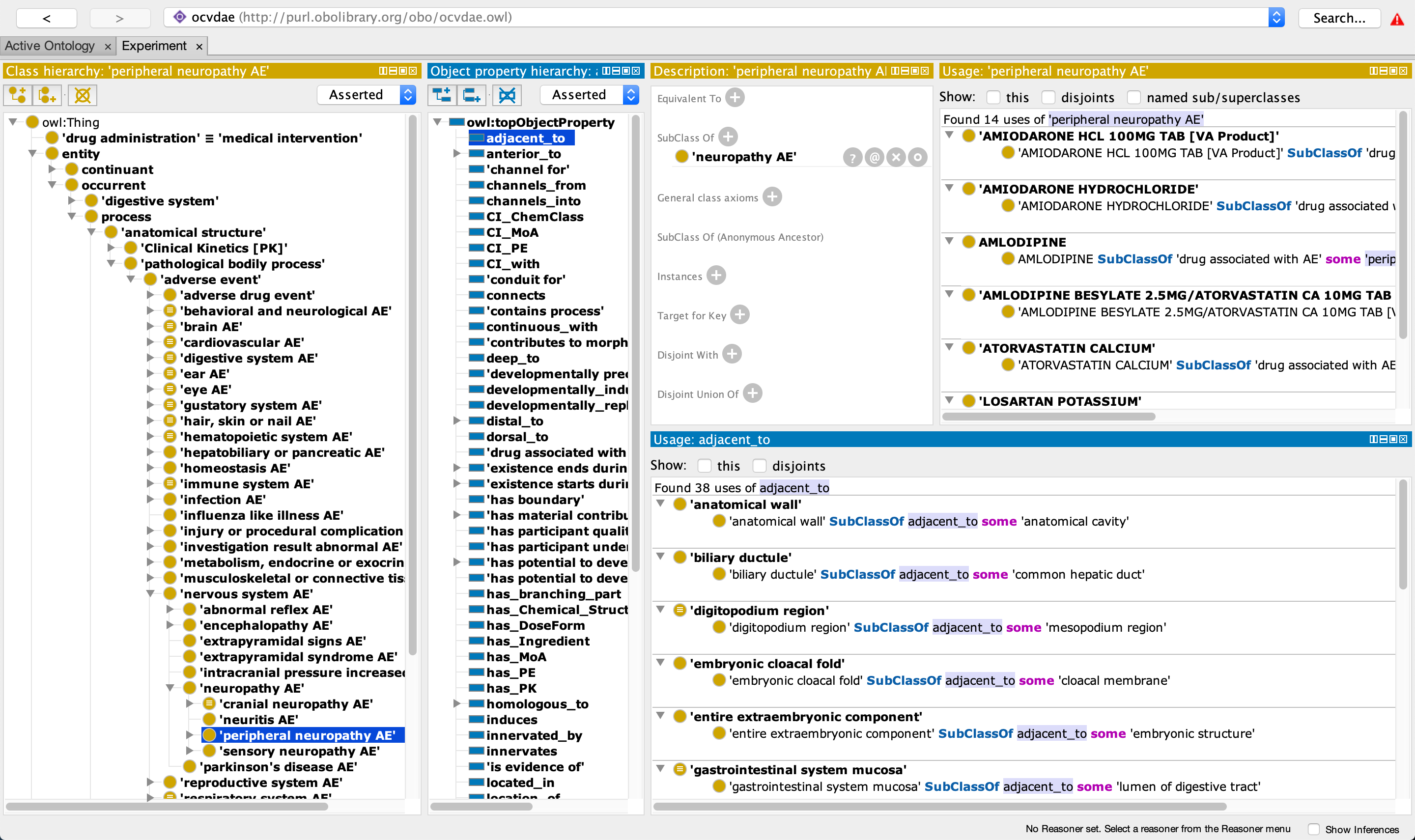}
  \caption{\protege interface as configured for use in the study. Left: Class pane, centre: Object property pane, right-half-top-left: Class Description view, right-half-top-right: Class Usage view, 
  right-half-bottom: Property Usage view.}
  \label{fig:protege-configuration}
\end{figure}

We choose \protege because we want to compare \ontoplot to a robust tool. \protege is the most widely used and actively maintained tool for ontology creation and editing in the ontology engineering community (based on citations). It provides a baseline representation---an indented list---for ontology hierarchy browsing and visualises non-hierarchical associations as text lists in separate views (see Figure~\ref{fig:protege-configuration}).

% \begin{figure}[!htb]
%   \centering
%   \includegraphics[width=\columnwidth]{protege-default}
%   \caption{Default \protege interface.}
%   \label{fig:protege-default}
%   \yf{Remove from the paper, and move to the supplementary material, and mention it in the main paper. }
% \end{figure}

\begin{figure}[!htb]
  \centering
  \includegraphics[width=0.7\columnwidth]{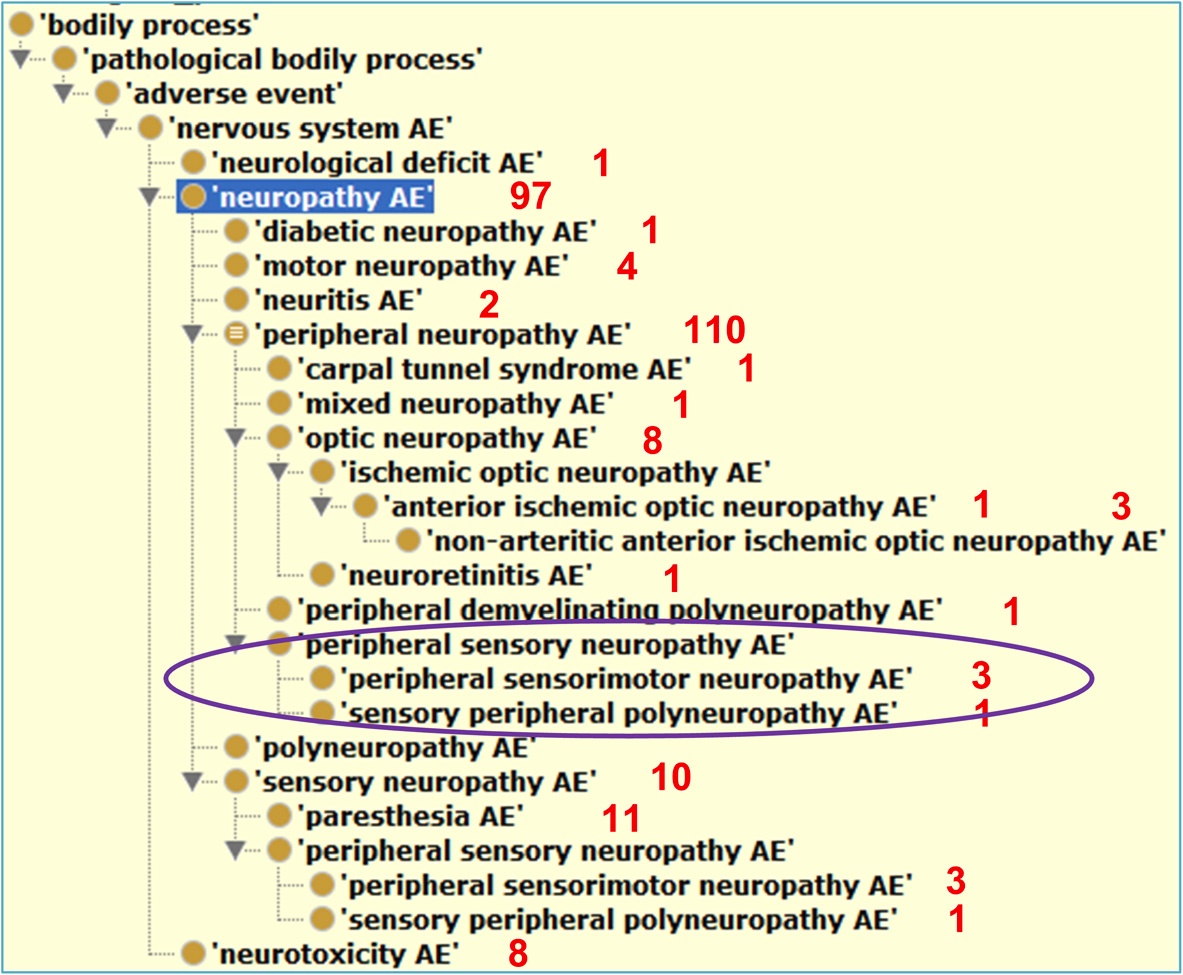}
  \caption{\hl{The manual approach} used by domain experts to show associations within an ontology: association numbers written next to branches of the hierarchy \hl{on a screenshot of} \protege indented list view (from\cite{guo2016ontology}, used with permission).}
  \label{fig:protege-label}
\end{figure}

Also, as mentioned in Section~\ref{sec:ont}, the domain experts we consulted (prior to the design of \ontoplot) frequently use \protege to perform their ontology-based analysis, and present their work using \hl{screenshots of} the \protege indented list view with manually added annotations to indicate the association strength in hierarchies~\cite{guo2016ontology}. (see Figure~\ref{fig:protege-label}).

%\mjw{We should get permission from Oliver to reproduce the image.} 
% * <yongqunhe@gmail.com> 2018-10-12T13:58:02.458Z:
% 
% You have got my permission now.  --Oliver
% 
% ^ <yongqunhe@gmail.com> 2018-10-12T13:58:39.031Z.

% While there have been several well-known plugins for \protege that support the display of the ontology hierarchy in a hierarchical structure and use a separate graph to visualise non-hierarchical associations (e.g., Jambalaya~\cite{storey2001jambalaya}, and Knoocks~\cite{jurcik2012knoocks}), to the best of our knowledge none of these tools are available any more and are no longer maintained. 

\protege is a fully-featured ontology engineering environment, and there are many panes, views and functionality not necessary for our experiment. To avoid confusing our participants with a complex interface, we simplified \protege by removing the unnecessary items from the interface, such as ``Data properties'' and ``Individuals'' panes, and the ``Class Annotations'' views. We also deselected some check boxes in the views and search window to avoid irrelevant information being shown to participants. To better support the tasks, we modified the interface layout of \protege to avoid view switching. We positioned the ``Object properties'' pane and the ``Classes'' pane side-by-side, placed the ``Class Description'' view and the ``Class Usage'' view next to each other on top of the ``Property Usage'' view. Figure~\ref{fig:protege-configuration} shows the interface layout configured to clearly show all the views and functions needed in the experiment. 

% Another popular tool, VOWL~\cite{lohmann2014vowl}, uses graph-based visualisation to model concept interrelations in ontologies, but does not differentiate hierarchical relationships and non-hierarchical associations visually. 
%As discussed in Section~\ref{sec:ont_vis}, \protege visualisation plugins---like the graph-based OWLViz~\cite{horridge2005owlviz} and WebVOWL~\cite{lohmann2014vowl}---were not used in our experiment due to either being unavailable or suffering from scalability issues with our chosen ontologies.
%\yf{Add a paragraph here about the tools we tried to evaluate against but can't (WebVOWL, various \protege plugins, etc.)}

\subsection{Tasks} % (fold)
\label{sub:tasks}

As mentioned in Section~\ref{sub:design_motivation}, we identified a range of important use cases and user needs for biomedical ontologies from the literature as well as from discussions with domain experts. To test the usability of \ontoplot with respect to the identified user needs, we designed ten tasks from the use cases and organised them into three groups, shown in Table~\ref{table:tasks}. 

%\begin{table*}[!htb]
%\centering
%\caption{Tasks in the experiment}
%\label{table:tasks}
%\begin{tabular}{ >{\arraybackslash}p{0.6in} | >{\arraybackslash}p{0.4in} | >{\arraybackslash}p{0.5in} | >{\arraybackslash}p{4.8in}}
 %\textbf{Group} & \textbf{Task} & \textbf{Use Case} & \textbf{Description}\\
%\hline
%Hierarchy & T1 &  & Identify the parent of a class.\\
 %& T2 &  & Identify the child(ren) of a class.\\
 %& T3 &  & Identify the sibling(s) of a class.\\
 %& T4 & U1 & Identify the path from a class to the root.\\
 %& T5 & U2 & Identify the closest common ancestor of two classes.\\ 
 %\hline
%Association & T6 &  & Identify the classes associated with a class.\\
 %& T7 &  & Identify the number of associations of a class.\\
 %& T8 & U3 & Identify the class having the highest number of associations.\\ 
 %\hline
%Hierarchy + & T9 & U4 & Identify the parent class with the most children who are associated with a class. \\
%Association & T10 & U5 & Identify a class that is not associated with a class, but all of its sibling(s) are associated with that class. \tcz{Should that be: Identify a class that is not associated with a class but all of its sibling(s) are associated with that class.}\\
%\end{tabular}
%\end{table*}

\begin{table*}[!htb]
\caption{Tasks in the experiment.}
\label{table:tasks}
\scriptsize%
\centering%
\begin{tabular}{lllll}
\toprule
 Group & Task & Use case & Description & \hl{Example instruction}\\
\midrule
G1. Hierarchy & T1 & U1 & Identify the parent of a class. & \hl{Please tell me the parent of ``skin of body''.}\\
 & T2 & U1 & Identify the child(ren) of a class. & \hl{Please tell me the children of ``limb segment''.}\\
 & T3 & U1 & Identify the sibling(s) of a class. & \hl{Please tell me the siblings of ``anatomical space''.}\\
 & T4 & U2 & Identify the path from a class to the root. & \hl{Please tell me the path from ``process'' to the root.}\\
 & T5 & U3 & Identify the closest common ancestor of & \hl{Please tell me the closest common ancestor of} \\ 
 & & & two classes. & \hl{``anatomical collection'' and ``anatomical surface''.} \\
 \midrule
G2. Association & T6 & U4 & Identify the classes associated with a class. & \hl{Please tell me the classes which have the ``may\_prevent''} \\
& & & & \hl{association with the ``Pain'' class.} \\
 & T7 & U4 & Identify the number of associations of a class. & \hl{Please tell me the number of ``may\_prevent'' associations} \\
 & & & & \hl{of the ``Hypertrophy'' class.} \\
 & T8 & U5 & Identify the class having the highest number & \hl{Please tell me the class which has the most ``may\_treat''} \\ 
 & & & of associations. & \hl{associations.} \\
 \midrule
G3. Hierarchy + & T9 & U6 & Identify the parent class with the most children & \hl{Please tell me the class which has the most children that have} \\
Association & & & who are associated with a class. & \hl{the ``adjacent\_to'' association with the ``full formed stage'' class.} \\
& T10 & U7 & Identify a class that is not associated with a & \hl{Please tell me the class whose siblings all have the} \\
& & & specified class, but all of its sibling(s) are & \hl{``site\_of\_metabolism'' association with the ``Channelopathy''} \\
& & & associated with that class. & \hl{class, but that class itself does not have such an association.} \\
\bottomrule
\end{tabular}
\end{table*}

For the first group of tasks (G1), we focus on the hierarchical structure of ontologies. \hl{While these are basic hierarchy comprehension tasks, they are essential to almost all analysis of ontologies. For example,  T1, T2, and T3 ask about parent-child relationships, requiring exploration of the ontology hierarchical structure (U1), and investigate whether the visual compression and glyphs in \ontoplot impact the cognition of the ontology hierarchy. Similarly, T4 asks a user to trace the hierarchical path from a class to the root which is related to generalising concepts (U2). T5 examines the intersection of two subtrees supporting common knowledge discovery (U3).}

%These tasks require a basic exploration of ontology hierarchical structure (T1, T2, T3), or are related to primary needs of generalising concepts (T4) or common knowledge discovery (T5). 

The second group of tasks (G2) focus on non-hierarchical associations. \hl{Both T6 and T7 require an exploration of the associations for a class (U4). While T6 asks for all classes associated with a class, T7 asks for the total number of them. T8 requires users to find the class with the highest number of associations in the ontology, which identifies significant classes (U5).}

%These tasks aim at investigating individual associations (T6, T7), or detecting significant classes (T8). 

The third group of tasks (G3) further examines the associations together with the hierarchical structure. These tasks are the most complex ones but essential for \hl{analysing associations on the class level. T9 asks for the parent having the most children with associations, which helps determining the \emph{class effect} (U6). T10 finds the outlier (class without associations) among a group of sibling classes with associations, providing evidence for predicting undiscovered associations (U7).}

%determining the ``class effect'' of associations (T9) or to provide evidence for predicting undiscovered associations (T10).

% subsection tasks (end)

\subsection{Hypotheses} % (fold)
\label{sub:hypothesis}
%\yf{Need to refer back to these hyphothesis in the discussion}

We hypothesised that \ontoplot would perform similarly to \protege for G1 hierarchy tasks (H1), since both tools clearly emphasise the hierarchical structure of ontologies. We believed that \ontoplot would outperform \protege on G2 association tasks (H2) and G3 hierarchy and association combined tasks (H3), since \ontoplot was designed to support ontology association analysis. 

% subsection hypothesis (end)

\subsection{Datasets} % (fold)
\label{sub:datasets}

We use two biomedical ontologies: CVDO~\cite{barton2014cardiovascular} and OCVDAE~\cite{wang2017ontology}. CVDO (536 classes) was used for the training tasks. OCVDAE (4,589 classes) was used for the study tasks. In total, there are 8 object properties and 551 non-hierarchical associations in CVDO. In OCVDAE, there are 118 object properties and 20,269 non-hierarchical associations. In order to keep the experiment to a reasonable time, we selected classes with less than 25 associations to ask questions about.  

The preliminary user study used a small manually constructed (and hence unrealistic) ontology for the training, and participants did the tasks with both CVDO and OCVDAE.  That study found little difference in the results between the two ontology sizes, hence the decision to evaluate only the larger ontology in this expert study and use the smaller ontology for the training tasks.

% \yy{In addition, we created a small training ontology with 15 classes, 2 object properties and 6 non-hierarchical associations to introduce the tasks to the participants, which covered all possible situations in the study.}

%\tcz{Is this important to be mentioned since we finally didn't use the training ontology?} \yy{I think this is a part of the story why we end up with only one size ontology for the actual tasks. I made it shorter.}
% Was 23 

% In addition, we created a training ontology to introduce the tasks to the participants. As most participants were not expected to have experience with ontologies, we kept the training ontology simple and small. The training ontology contains 15 classes, 2 object properties, and 6 associations, but covers all possible situations in the larger ontologies used in the study. 
%For example, the examined siblings could be separated in different boxes, and the examined children could be initially hidden in glyphs.  \mjw{This last sentence doesn't make sense to me.} \tcz{For me the last two sentences need some rephrasing.} 

% subsection datasets (end)

\subsection{Procedure} % (fold)
\label{sub:procedure}

We used a within-subjects design for the experiment: $2\text{ tools}\times 1\text{ ontology size}\times 10\text{ tasks } (+\text{ training})$. 

%Participants performed tasks using the same ontology with different tools. To avoid issues of memorisation, we systematically renamed and shortened all class and object property labels to be different when used in each tool.

To ensure consistent difficulty of tasks, the same ontology was used for the tasks performed using each tool.  To avoid issues of memorisation, class and object property labels were consistently renamed to be different for each tool.

We fixed the order of tasks for each tool but counterbalanced the order of tools shown to different participants.

Participants were required to complete training before performing the experimental tasks. They were firstly shown an introductory document to explain the terminology used in the experiment.
% \tcz{Ying, please revise as required.}. 
Participants also finished a training for each tool before using them. They were shown an introductory document to demonstrate the interface and functions of the tool and were then required to use the tool to answer 10 sample questions with the training ontology.
% \tcz{Ying, please revise as required.}. 
The sample questions covered all experiment tasks in order to allow participants to be familiarised with the tools and the tasks. While answering the sample questions, participants were guided to practise the functions that were needed in the actual tasks for each tool, such as searching, clicking classes or object properties, double-clicking to expand or collapse subtrees, hovering the mouse cursor over classes to read class labels and association information, and marking classes by pinning labels on them in \ontoplot, 
% \tcz{Ying, please revise as required.}, 
and going back or forward in \protege. After each question, participants were shown the correct answer, and an explanation was given if they didn't answer correctly.  

%\yf{Do we still use a web site to collect answers?}

%A website has been developed to collect participants' answers and to record completion time for each task. When participants were ready to begin a task, they clicked a button to load that task. When they finished a task, they clicked a button to indicate they had completed this task and were ready to progress to the next task. As switching object properties in OCVDAE in \ontoplot required a couple of seconds to load the visualisation, we excluded the loading time from the task completion time. To keep the experiment within a reasonable time, we set up a time limit for each task of 2 minutes. For any task, if participants found it too difficult to answer, they could choose to skip that task.

The participants were given access to a study website that guided them through the study, gave them access to training instructions, tasks, and survey questions.
When the participants started a task, this was recorded by the investigator. When they completed a task, the participant would signal this to the investigator who would record their answer and completion time.   For any task, if participants found it too difficult to complete, they could choose to skip that task.

After completing the tasks for each tool, participants were asked to complete a survey, rating the difficulty level and the confidence level of their answers for each group of tasks. We also collected participants' preferences and comments at the end of the experiment. Answers to survey questions were entered by participants into a Google Form.
% As \protege is an existing system, we also asked participants whether they had used \protege before the experiment. 

After the experiment, participants were asked to answer some questions regarding their background knowledge and experience with ontologies, \protege, and ontology visualisation tools.

Each experiment session lasted approximately one and a half hours, including training and surveys. 

\subsection{Participants and Apparatus}\label{sub:participants_and_apparatus_2}

All 12 participants had experience in the field of ontologies or knowledge graphs. Eleven of them identified as having experience using ontologies, including three with more than three years experience. 
Ten participants had experiences using \protege, one of whom had more than three years of experience. 
Another two participants had used other ontology tools, including the tools developed by the Gene Ontology Consortium and a proprietary tool used for a knowledge graph construction engine. 
Of the 12 participants, three were female and nine were male. Their age ranged from 18 to 41. 
All participants had normal or corrected-to-normal vision, and none suffered colour blindness.

The six participants recruited from the authors' university used a 2.3 GHz Intel Core i5 laptop with 8GB of RAM, using a 24-inch monitor with a resolution of 3840x2160 pixels. 
The six participants recruited from other institutions did the experiment remotely, using their own computers at a resolution of 1600x900 pixels. 
For the remote participants the experiments were observed via video call.

\subsection{Results}\label{sub:results_2}
%\yf{Added the 2nd sentence below on study design. Please see if it's repetitive/enough.}

All 12 participants completed the study. Unlike in the preliminary study (prototype evaluation described in Section~\ref{sec:user_study_1}), all 12 participants completed all tasks on the same ontology, OCVDAE. 
We measured accuracy and completion time for each task, and collected difficulty level, confidence level, preference ranking, and learning effort as rated by the participants. As the data is not normally distributed, we used the non-parametric \textit{Wilcoxon test} to compare accuracy between the two tools~\cite{field2012discovering}.
% We used the \textit{Wilcoxon test} to analyse the accuracy and subjective rated data. 
% We again removed the completion time for accuracy that is not greater than 0\%, and used the \textit{Whitney-Mann test} for these unequal data. 
For the completion time data, we only considered the time for answers with an accuracy greater than 0\%. Therefore, we used the non-parametric \textit{Whitney-Mann test} for unequal samples~\cite{niroumand2013statistical}. For the rated results, we also used \textit{Wilcoxon test} to analyse significance.

The main evaluation results are summarised in Table~\ref{table:results_sum_s2}, including statistical significance for each item. Below we discuss them in detail.

% \begin{table*}[!htb]
% \caption{Summary of the second user study results (O: \ontoplot, P: \protege).}
% \label{table:results_sum_s2}
% \scriptsize%
% \centering%
% \begin{tabu}{llllllll}
% \toprule
%  Group & Task & Accuracy & Time & Difficulty & Confidence & Preference & Learning Effort \\
% \midrule
% \multirow{5}{*}{G1} & T1 & O $=$ P & O $>$ P & \multirow{5}{*}{O $>$ P} & \multirow{5}{*}{O $>$ P} & \multirow{5}{*}{O $<$ P} & \multirow{10}{*}{O $<$ P}\\
% & T2 & O $>$ P & O $>$ P ** & & & &\\
% & T3 & O $>$ P & O $>$ P ** & & & &\\
% & T4 & O $=$ P & O $>$ P & & & &\\
% & T5 & O $>$ P & O $<$ P & & & &\\
% \cmidrule{1-7}
% \multirow{3}{*}{G2} & T6 & O $>$ P & O $>$ P & \multirow{3}{*}{O $<$ P } & \multirow{3}{*}{O $>$ P} & \multirow{3}{*}{O $\gg$ P} &\\
% & T7 & O $>$ P & O $<$ P *** & & & &\\
% & T8 & O $>$ P * & O $<$ P *** & & & &\\
% \cmidrule{1-7}
% \multirow{2}{*}{G3} & T9 & O $>$ P & O $<$ P *** & \multirow{2}{*}{O $<$ P} & \multirow{2}{*}{O $>$ P} & \multirow{2}{*}{O $\gg$ P} &\\
% & T10 & O $>$ P & O $<$ P ** & & & &\\
% \midrule
% & & \multicolumn{6}{c}{***p $<$ 0.001\quad **p $<$ 0.01\quad *p $<$ 0.05}\\
% \bottomrule
% \end{tabu}
% \end{table*}

\begin{figure}[!htb]
  \centering
  \begin{subfigure}[b]{0.441\columnwidth}
    \centering
    \includegraphics[width=\columnwidth]{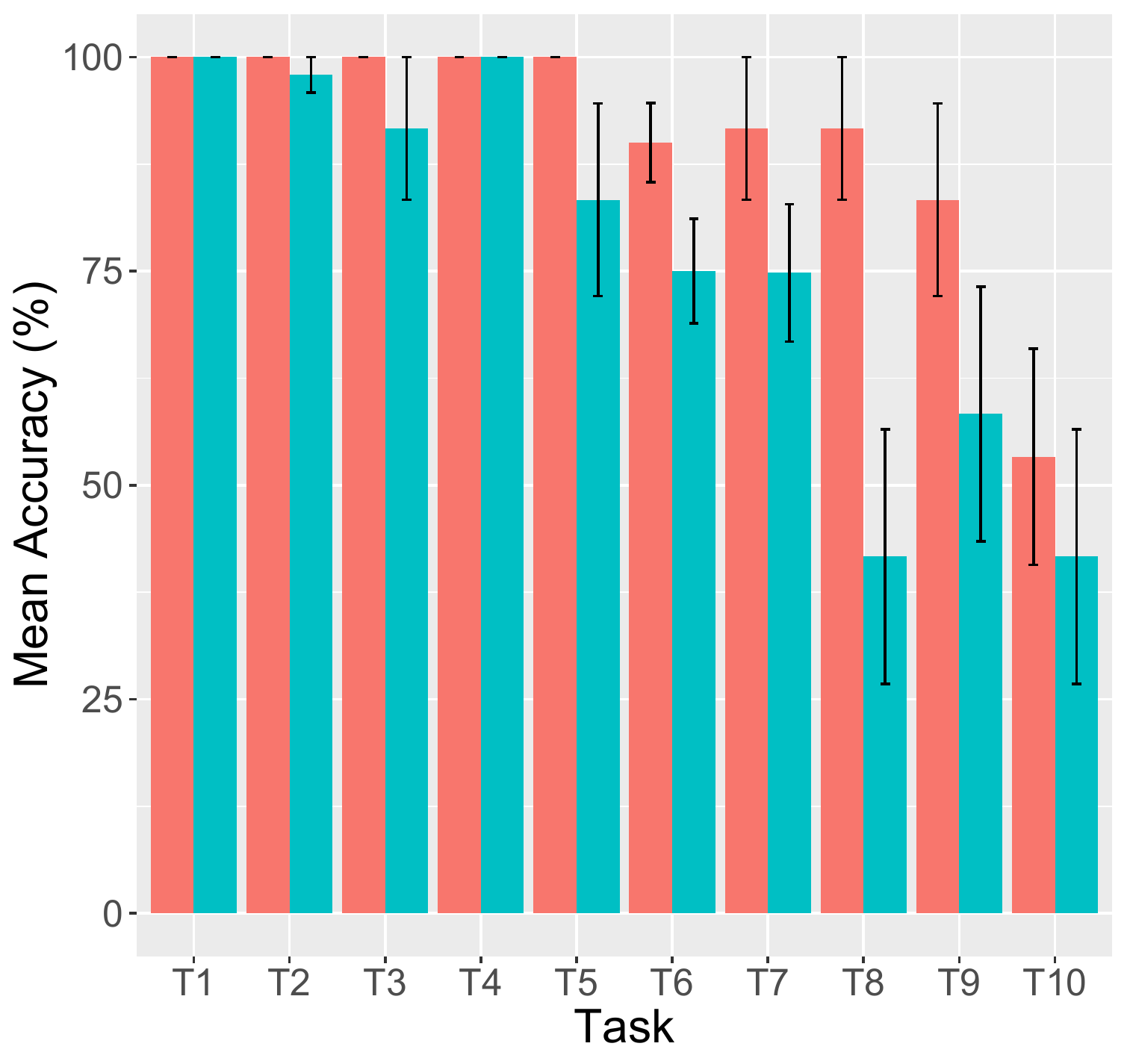}
    \caption{Mean accuracy for each tool per task.}
    \label{fig:accuracy_s2}
  \end{subfigure}
  \hfill
  \begin{subfigure}[b]{0.55\columnwidth}
    \centering
    \includegraphics[width=\columnwidth]{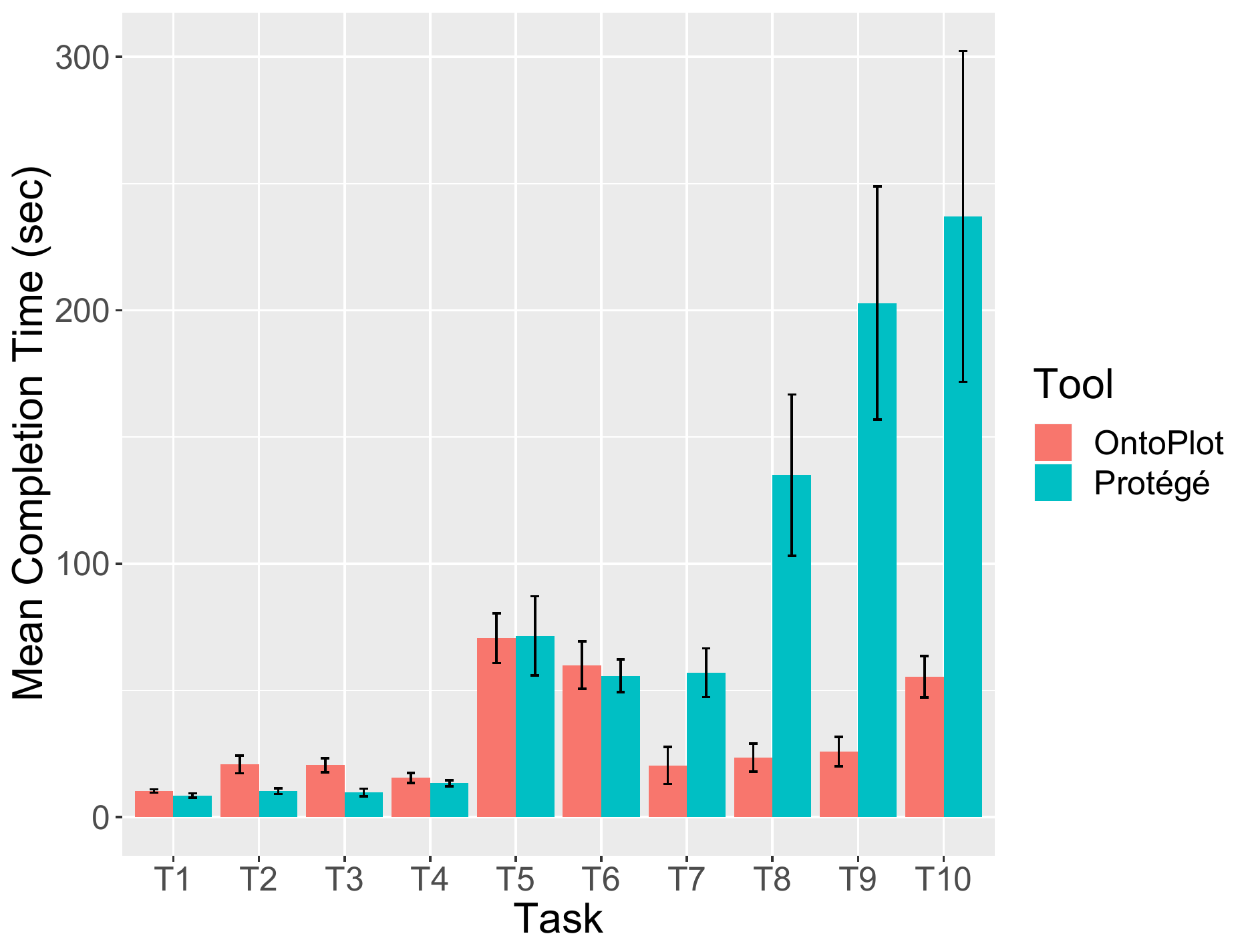}
    \caption{Mean completion time for each tool per task.}
    \label{fig:time_s2}
  \end{subfigure}
  
  \caption{Participants' performance of the two tools in the \hl{expert} user study.}
\end{figure}

\textbf{Accuracy.} Figure~\ref{fig:accuracy_s2} shows the details of mean accuracy for each tool per task. We found overall, participants achieved higher accuracy on most tasks with \ontoplot than with \protege. The two exceptions are for T1 (finding parent) and T4 (finding path), which have equal accuracy (100\%) for both tools. The \textit{Wilcoxon test} revealed that for T8 (finding class with most associations), \ontoplot significantly outperformed \protege (p $<$ 0.05). 

% \yf{What 'significantly' mean here? 'substantially', 'statistically significantly' or both? If both, then we need to rephrase it.} \yy{I added \textit{Wilcoxon test} at the beginning of this sentence.}

% \yf{From Table 7 isn't it T9 and T10 are also statistically significant?}

% \begin{figure}[!htb]
%   \centering
%   \includegraphics[width=0.7\columnwidth]{accuracy_s2}
%   \caption{Mean accuracy for each tool per task.}
%   \label{fig:accuracy_s2}
% \end{figure}
 
\textbf{Completion Time.} Results for completion time are shown in Figure~\ref{fig:time_s2}. We found for most of the Hierarchy tasks (G1), participants spent less time on \protege than on \ontoplot. Especially for T2 (finding children) and T3 (finding siblings), The \textit{Whitney-Mann test} revealed that \protege significantly outperformed \ontoplot (p $<$ 0.01). For T5 (finding common ancestor), \ontoplot and \protege had very close completion time, with \ontoplot being slightly faster. Of the Association tasks (G2), for task T6 (finding individual associations) \ontoplot had a slightly longer completion time than \protege. The results show that tasks for finding and counting most associations (T7, T8), \ontoplot significantly outperformed \protege (p $<$ 0.001). Highly significant differences were also found for the combined Hierarchy + Association tasks (G3) (T9, T10), with \ontoplot substantially outperforming \protege. 
Taking accuracy into account, these results indicate that, especially for complex tasks (G3), \ontoplot requires substantially less time and achieves much higher accuracy than \protege.

% \begin{figure}[!htb]
%   \centering
%   \includegraphics[width=0.7\columnwidth]{time_s2}
%   \caption{Mean completion time for each tool per task.}
%   \label{fig:time_s2}
% \end{figure}

\begin{figure}[!htb]
  \centering
  \begin{subfigure}[b]{0.57\linewidth}
    \includegraphics[width=\textwidth]{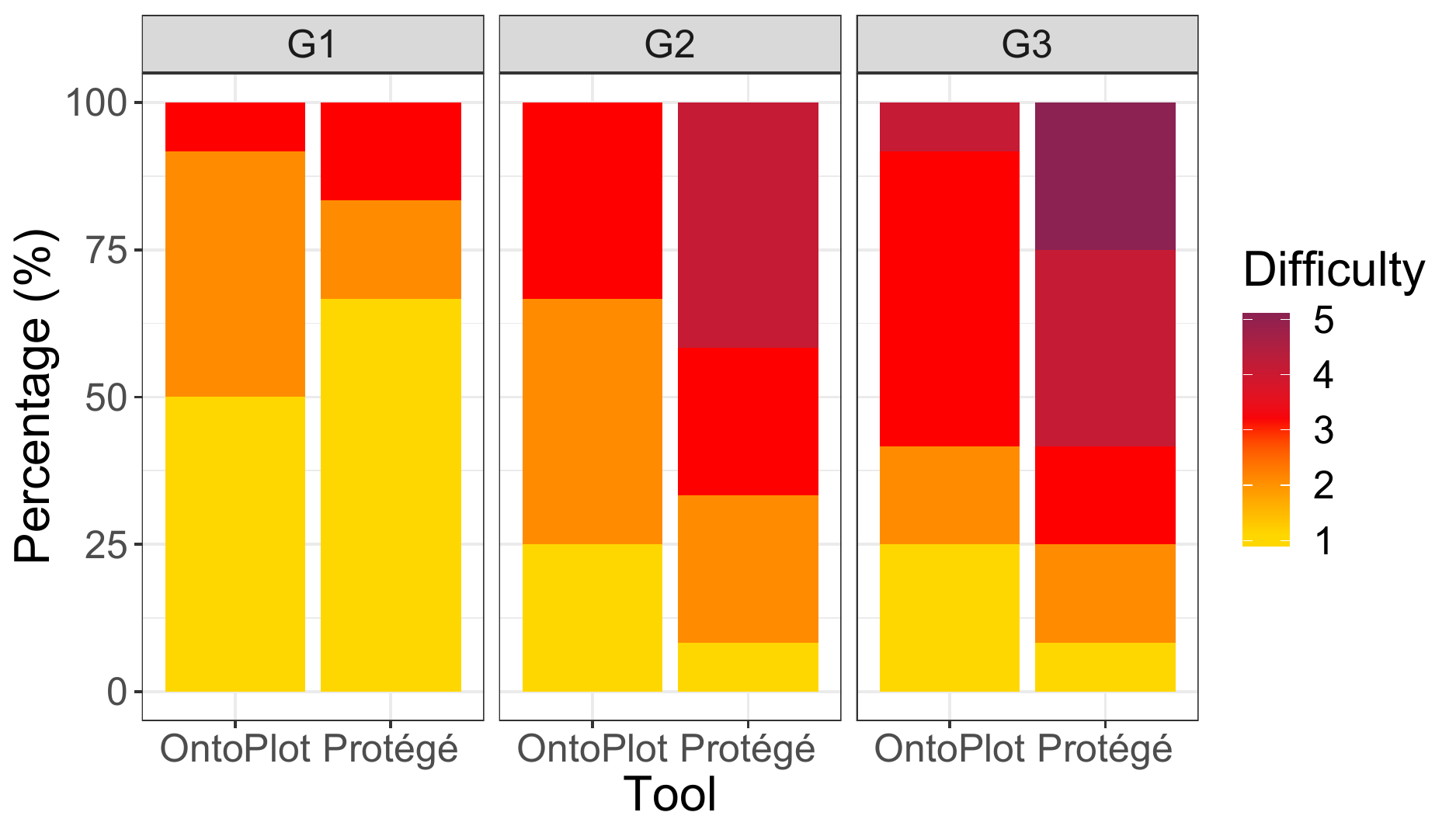}
    \caption{\hl{Difficulty rating for task groups.}}
    \label{fig:difficulty_s2}
  \end{subfigure}
  \hfill
  \begin{subfigure}[b]{0.42\linewidth}
    \includegraphics[width=\textwidth]{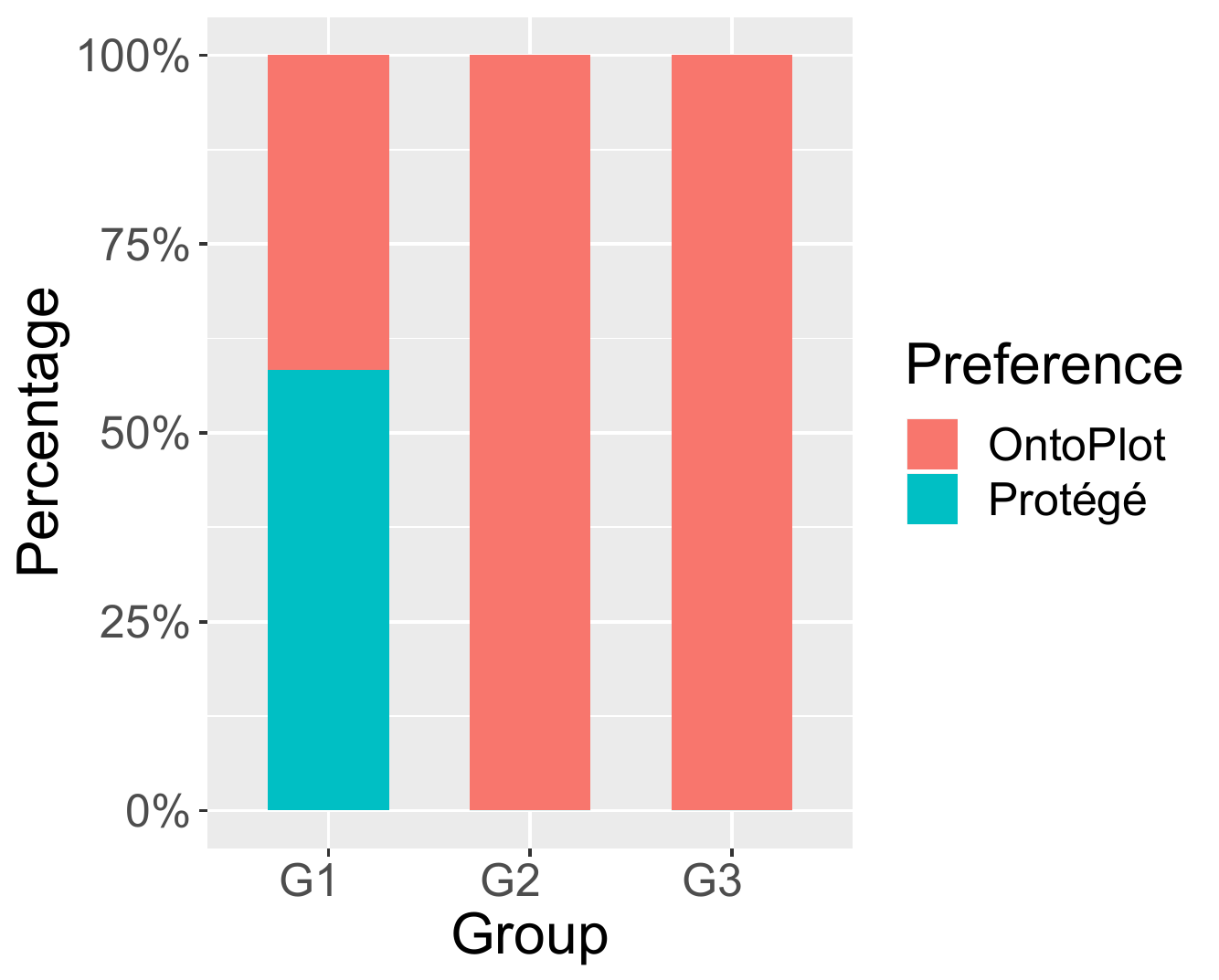}
    \caption{\hl{Preference rating for task groups.}}
    \label{fig:preference_s2}
  \end{subfigure}
  
  \begin{subfigure}[b]{0.57\linewidth}
    \includegraphics[width=\textwidth]{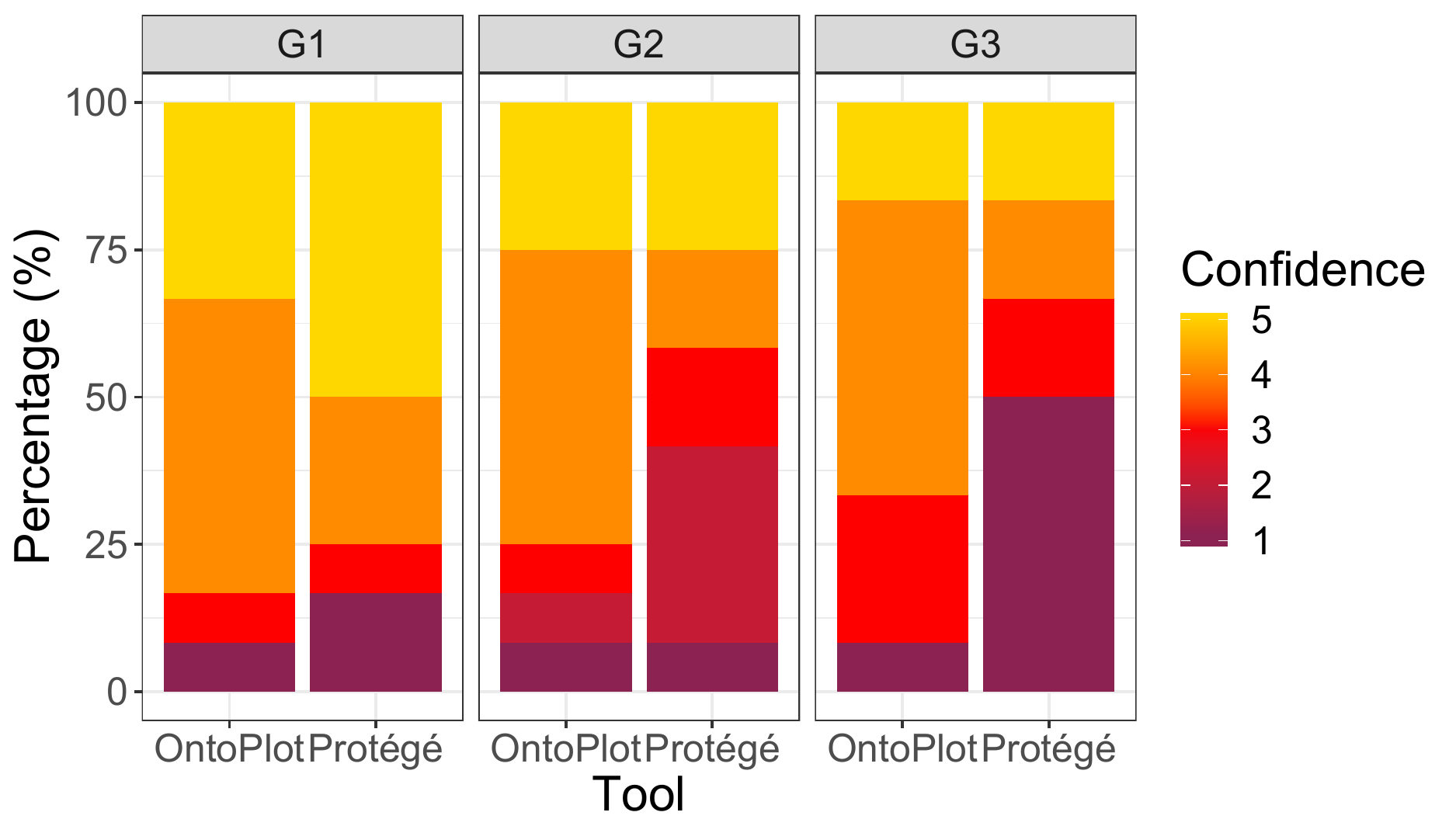}
    \caption{\hl{Confidence rating for task groups.}}
    \label{fig:confidence_s2}
  \end{subfigure}
  \hfill
  \begin{subfigure}[b]{0.42\linewidth}
    \includegraphics[width=\textwidth]{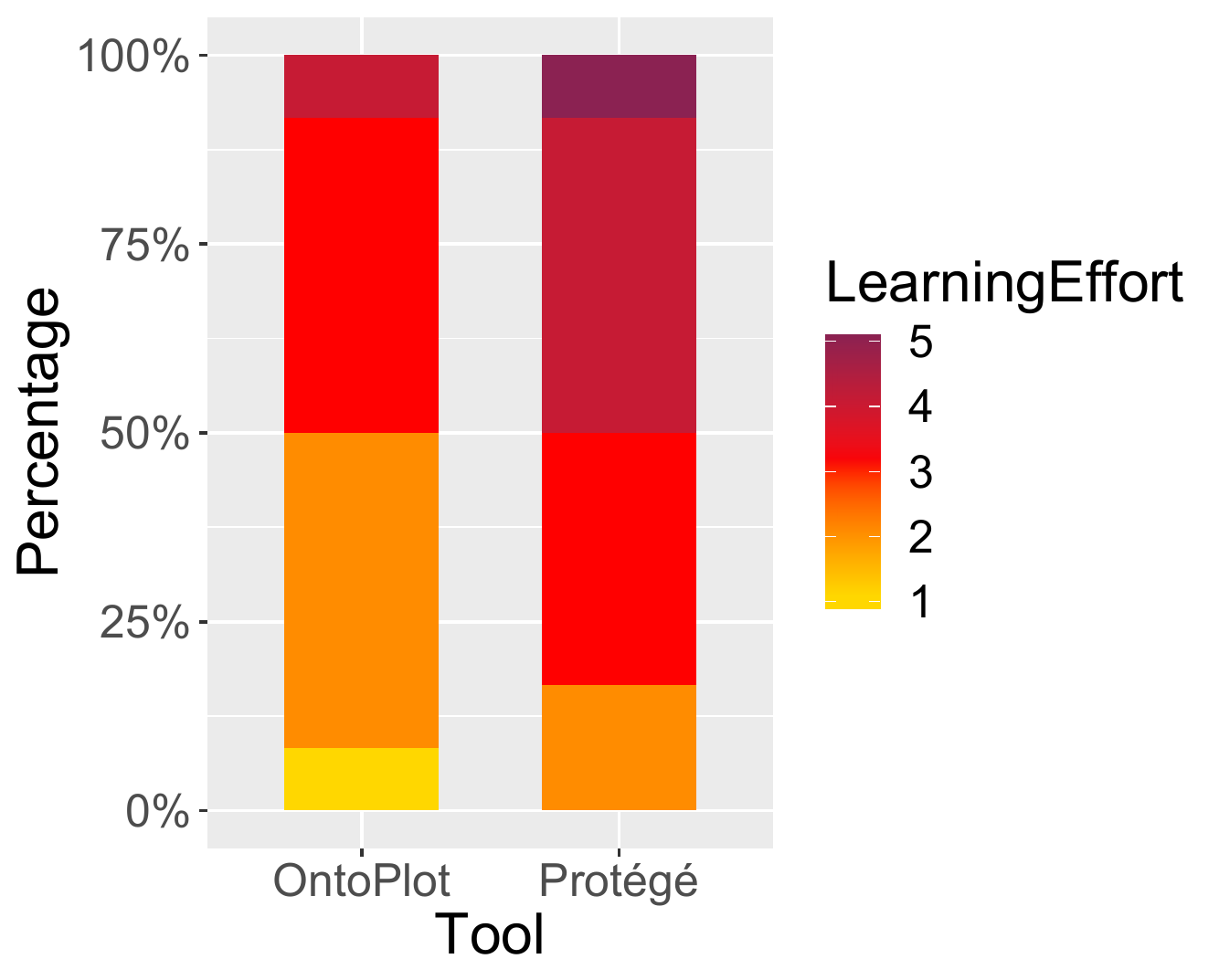}
    \caption{\hl{Learning effort rating.}}
    \label{fig:learning-effort_s2}
  \end{subfigure}

\caption{Participants' rating of the two tools in the \hl{expert} user study.}
\label{fig:rating_s2}
\end{figure}

\textbf{Participant Rating.} We use a five-point Likert scale ranging from 1--5 to measure participants' rating of difficulty (lower is better) and confidence (higher is better) for each group of tasks and each tool. Figure~\ref{fig:difficulty_s2} and Figure~\ref{fig:confidence_s2} show the percentage of participants' rating results, and Table~\ref{table:diff_conf} summarises the results.

\begin{table}[!htb]
\caption{Summary of average of difficulty and confidence ratings as shown in Figures~\ref{fig:difficulty_s2} and \ref{fig:confidence_s2}.}
\label{table:diff_conf}
\scriptsize%
\centering%
\begin{tabu}{lllll}
\toprule
\multirow{2}{*}{Group} & \multicolumn{2}{c}{Difficulty} & \multicolumn{2}{c}{Confidence}\\
\cmidrule{2-5}
& \ontoplot & \protege & \ontoplot & \protege \\
\midrule
G1 & 1.583 & 1.5 & 4 & 3.917\\
G2 & 2.083 & 3 & 3.75 & 3.167\\
G3 & 2.417 & 3.5 & 3.667 & 2.5\\
% \midrule
% & \multicolumn{2}{c}{***p $<$ 0.001 **p $<$ 0.01 *p $<$ 0.05}\\
\bottomrule
\end{tabu}
\end{table}

Overall, participants rated G1 tasks performed in \protege as slightly less difficult than in \ontoplot. For G2 and G3 tasks, participants rated \ontoplot as less difficult than \protege. Three participants rated \protege difficulty at 5 (highest) for G3 tasks. 

When asked about confidence rating, participants felt slightly more confident with \ontoplot than with \protege for G1 tasks and gave much higher confidence rating to \ontoplot for G2 and G3 tasks. 

Figure~\ref{fig:preference_s2} shows the result of the preference rating. For G1 tasks, seven participants preferred \protege over \ontoplot, whereas the situation is entirely reversed for G2 and G3 tasks. All the participants preferred \ontoplot for these tasks.

The result of the learning effort rating is shown in Figure~\ref{fig:learning-effort_s2}, also using a five-point Likert scale ranging from 1 (easiest) to 5 (hardest). One participant rated learning effort 1 for \ontoplot, while one participant rated it 5 for \protege. The average rating is 2.625 for \ontoplot and 3.25 for \protege. There is no significant difference between the tools (p = 0.056).

% \begin{figure*}[!htb]
%   \centering
%   \begin{subfigure}[b]{0.27\linewidth}
%     \includegraphics[width=\textwidth]{difficulty_s2}
%     \caption{}
%     \label{fig:difficulty_s2}
%   \end{subfigure}
%   \begin{subfigure}[b]{0.27\linewidth}
%     \includegraphics[width=\textwidth]{confidence_s2}
%     \caption{}
%     \label{fig:confidence_s2}
%   \end{subfigure}
%   \begin{subfigure}[b]{0.222\linewidth}
%     \includegraphics[width=\textwidth]{preference_s2}
%     \caption{}
%     \label{fig:preference_s2}
%   \end{subfigure}
%   \begin{subfigure}[b]{0.19\linewidth}
%     \includegraphics[width=\textwidth]{learning-effort_s2}
%     \caption{}
%     \label{fig:learning-effort_s2}
%   \end{subfigure}

% \caption{Participants' rating of the two tools in the second user study: (a) difficulty rating for each group of tasks, (b) confidence rating for each group of tasks, (c) preference rating for each group of tasks, and (d) learning effort rating.}
% \label{fig:rating_s2}
% \end{figure*}

\textbf{Participant Feedback.} At the end of the experiment each participant was given the chance to provide feedback and give comments. Some participants felt \protege was more familiar and acceptable, e.g., commenting ``The vertical aligned indented list is easier to perceive hierarchy structures''. Most of the participants gave positive feedback for \ontoplot, e.g., commenting ``\ontoplot interface is more friendly'', ``\ontoplot needs effort to learn, but makes tasks easier'', or ``\ontoplot has more compact view of the ontology''. Some participants also provided more specific feedback such as ``Lighter lines and darker lines are helpful for distinguishing siblings and non-siblings'', ``The labels make finding associations much easier'', ``Association labels are easy to read'', or ``Tagging feature is nice''. One participant also commented on \protege: ``That is very difficult to find common ancestors with \protege''.

A few participants also provided helpful feedback for further improvements of \ontoplot, e.g., ``Probably can use colour coding for the sibling lines to make them more obvious'', ``The subtle arrows could be more effective if can indicate the number of pointing in and pointing out associations'', or ``Probably can filter association classes further when there are many associations''.

\textbf{Summary.} 
Table~\ref{table:results_sum_s2} presents a summary of all the results. 
Overall, \ontoplot moderately outperformed \protege on accuracy for \hl{most} tasks, and significantly \hl{(i.e., statistically significantly)} outperformed \protege for the task T8. On completion time, \ontoplot was outperformed by \protege for most G1 tasks \hl{(significantly for two tasks)}, while \ontoplot significantly outperformed \protege for most G2 and G3 tasks. No significant difference was revealed by the statistical test for the participants' rating data. 
These results are consistent with those from the first user study, while in the expert study the users had noticeably better accuracy rates using both tools and they performed significantly faster using \ontoplot than \protege on the G2 and G3 (association) tasks.

%\yy{TODO: improve readability}

% \begin{table*}[!htb]
% \caption{Summary of the second user study results (O: \ontoplot, P: \protege).}
% \label{table:results_sum_s2}
% \scriptsize%
% \centering%
% \begin{tabu}{llllllll}
% \toprule
%  Group & Task & Accuracy & Time & Difficulty & Confidence & Preference & Learning Effort \\
% \midrule
% \multirow{5}{*}{G1} & T1 & O $=$ P & O $>$ P & \multirow{5}{*}{O $>$ P} & \multirow{5}{*}{O $>$ P} & \multirow{5}{*}{O $<$ P} & \multirow{10}{*}{O $<$ P}\\
% & T2 & O $>$ P & O $>$ P ** & & & &\\
% & T3 & O $>$ P & O $>$ P ** & & & &\\
% & T4 & O $=$ P & O $>$ P & & & &\\
% & T5 & O $>$ P & O $<$ P & & & &\\
% \cmidrule{1-7}
% \multirow{3}{*}{G2} & T6 & O $>$ P & O $>$ P & \multirow{3}{*}{O $<$ P } & \multirow{3}{*}{O $>$ P} & \multirow{3}{*}{O $\gg$ P} &\\
% & T7 & O $>$ P & O $<$ P *** & & & &\\
% & T8 & O $>$ P * & O $<$ P *** & & & &\\
% \cmidrule{1-7}
% \multirow{2}{*}{G3} & T9 & O $>$ P & O $<$ P *** & \multirow{2}{*}{O $<$ P} & \multirow{2}{*}{O $>$ P} & \multirow{2}{*}{O $\gg$ P} &\\
% & T10 & O $>$ P & O $<$ P ** & & & &\\
% \midrule
% & & \multicolumn{6}{c}{***p $<$ 0.001\quad **p $<$ 0.01\quad *p $<$ 0.05}\\
% \bottomrule
% \end{tabu}
% \end{table*}

\begin{table}[!htb]
\caption{Summary of the \hl{expert} user study results \hl{(Acc: Accuracy, Diff: Difficulty, Conf: Confidence, Pref: Preference, L.Effort: Learning Effort). The tool mentioned in the columns outperforms the other in terms of the given metric} (O: \ontoplot, P: \protege).}
\label{table:results_sum_s2}
\scriptsize%
\centering%
\begin{tabu}{llllllll}
\toprule
 Group & Task & Acc & Time & Diff & Conf & Pref & L.Effort \\
\midrule
\multirow{5}{*}{G1} & T1 & - & P & \multirow{5}{*}{P} & \multirow{5}{*}{O} & \multirow{5}{*}{P} & \multirow{10}{*}{O}\\
& T2 & O & P** & & & &\\
& T3 & O & P** & & & &\\
& T4 & - & P & & & &\\
& T5 & O & O & & & &\\
\cmidrule{1-7}
\multirow{3}{*}{G2} & T6 & O & P & \multirow{3}{*}{O} & \multirow{3}{*}{O} & \multirow{3}{*}{O} &\\
& T7 & O & O*** & & & &\\
& T8 & O* & O*** & & & &\\
\cmidrule{1-7}
\multirow{2}{*}{G3} & T9 & O & O*** & \multirow{2}{*}{O} & \multirow{2}{*}{O} & \multirow{2}{*}{O} &\\
& T10 & O & O** & & & &\\
\midrule
\multicolumn{8}{c}{Significance: ***p $<$ 0.001\quad **p $<$ 0.01\quad *p $<$ 0.05}\\
\bottomrule
\end{tabu}
\end{table}

\subsection{Discussion}
%\yf{Talk about the hyphotheses in Section 4.3}

The expert user study shows that \ontoplot slightly outperformed \protege for Hierarchy tasks (G1) on accuracy, which accepted our hypothesis H1 (Section~\ref{sub:hypothesis}). A common error made by several participants in \protege was to mistake the sibling shown above a class (at the same indentation level) as the parent of that class, often when there was a some distance between them in the indented list. On completion time, \ontoplot was significantly outperformed by \protege for the tasks of finding children and siblings. This can be explained by the fact that most participants were \protege users and were familiar with the indented list for showing hierarchy structure. Also, in order to test the participants' perception of glyph compression, this group of tasks was designed to force participants to collapse or expand the subtrees. The participants spent some time on understanding which glyph or class they should collapse or expend in \ontoplot, and double-checked their answers. In \protege most of the participants can skilfully interact with the indented list. However, for the finding common ancestor task, the participants spent a little less time in \ontoplot than in \protege as they can mark the classes by labels, and this made the task easier.

For association-related tasks (G2 and G3), \ontoplot outperformed \protege on most of the tasks as expected (accepting H2, H3). Especially, for the completion time, there are some significant differences between the tools. We observed that the main reason why participants spent more time in \protege was because in \protege a user cannot select both classes and associations at the same time. Thus, the participants had to distinguish different classes or associations by themselves. We also observed that the reason why \ontoplot took marginally more time for task T6 (finding individual association classes) was that some participants spent some time on scrolling the visualisation or dragging the association labels.

\section{Conclusion and Future Work}\label{sec:conclusion}

%\todo{Add this.}

Expressive ontologies contain rich information captured by complex associations involving classes, properties and individuals. However, most existing ontology visualisation systems focus on class hierarchies, making it hard to find information about these associations. In this paper, we presented \ontoplot, a novel visualisation system specifically designed to support the interrogation of \emph{non-hierarchical} associations while still showing the class hierarchy of an ontology. \ontoplot improves space efficiency by a combination of hybrid icicle plots, visual compression techniques and interactivity. We compared \ontoplot with \protege, the de facto ontology editor, and found that \ontoplot significantly outperformed \protege on efficiency for the complex association-based tasks and was strongly favoured by the domain experts.
\hl{While we have evaluated \ontoplot on ontologies, it can be applied to other hierarchically structured data, e.g., research collaborations between organisations, where the number of relationships between individuals in the hierarchy shows how often the researchers worked together.}

% The results of the user study led us to think about how to further improve the usability of \ontoplot. 
We plan to add a mini map with linking and brushing to \ontoplot, for easier navigation and giving an overview of the entire ontology. \hl{Furthermore, we plan to improve the visual glyphs by providing more information about the size of hidden subtrees. As the same association may have different connectivities in different ontologies~\cite{burger2004integrating}, we will investigate how to represent multiple hierarchical structures with \ontoplot and show the differences between them in a clear manner.}

% , as well as user ratings, preference and confidence.

% Based on user feedback, we will further improve the usability and scalability of \ontoplot. 

% The authors wish to thank A, B, and C. This work was supported in part by
% a grant from XYZ (\# 12345-67890).}

%\clearpage

%\bibliographystyle{abbrv}
\bibliographystyle{abbrv-doi}

\bibliography{bibfile}
\end{document}